\documentclass[acmsmall]{acmart}
\usepackage{booktabs} 
\usepackage{subfiles}
\usepackage{balance}
\usepackage[utf8]{inputenc}
\usepackage{xcolor}
\setcopyright{acmlicensed}

\setcopyright{acmlicensed}
\acmJournal{TOCHI}
\acmYear{2020} \acmVolume{1} \acmNumber{1} \acmArticle{1} \acmMonth{1} \acmPrice{15.00}\acmDOI{10.1145/3386358}

\begin{document}
\title[Effects of Distraction on Smart Rings]{It’s All in the Timing: Principles of Transient Distraction Illustrated with Vibrotactile Tasks}

\author{Christopher L. Asplund}
\affiliation{%
  \institution{Division of Social Sciences, Yale-NUS College;}
  \institution{N.1 Institute for Health, National University of Singapore}
  \city{Singapore}
}
\email{chris.asplund@yale-nus.edu.sg}

\author{Takashi Obana}
\affiliation{%
  \institution{Division of Social Sciences, Yale-NUS College}
  \city{Singapore}
}
\email{takashi@yale-nus.edu.sg}

\author{Parag Bhatnagar}
\affiliation{%
  \institution{Division of Sciences, Yale-NUS College}
  \city{Singapore}
}
\email{paragbhtngr@gmail.com}

\author{Xun Quan Koh}
\affiliation{%
  \institution{Division of Social Sciences, Yale-NUS College}
  \city{Singapore}
}
\email{xunquan@yale-nus.edu.sg}

\author{Simon T. Perrault}
\affiliation{%
  \institution{Division of Sciences, Yale-NUS College;}
  \institution{Singapore University of Technology and Design (SUTD);}
  \city{Singapore}
  \institution{Korean Advanced Institute of Science and Technology (KAIST)}
  \city{Republic of Korea}
}
\email{perrault.simon@gmail.com}

%
%
%

\renewcommand{\shortauthors}{Asplund et al.}

\begin{abstract}
Vibration is an efficient way of conveying information from a device to its user, and it is increasingly used for wrist or finger-worn devices such as smart rings.
Unexpected vibrations or sounds from the environment may disrupt the perception of such information.
Although disruptive effects have been systematically explored in vision and audition, they have been less examined in the haptic domain.
Here we briefly review the relevant literature from HCI and psychology, distilling principles of when distraction is likely. We then investigate these principles through four experiments, examining how the timing and modality of relatively rare or unexpected stimuli (surprise distractors) affects the detection and recognition of vibrotactile target patterns.
At short distractor-target delays (< 350 ms), both auditory and vibrotactile surprise distractors impaired performance. At a longer delay (1050 ms), performance was not affected overall, even being improved with repeated exposure to the vibrotactile distractors.
We discuss the importance of our findings in the context of HCI and cognitive psychology, and we provide design guidelines for mitigating the effects of distraction on haptic devices.
\end{abstract}

%
%




\maketitle
\section{Introduction}

Wearable devices can convey information to users via different sensory modalities.
Visual and auditory interfaces are already used on numerous devices, including smart watches, activity trackers, and headsets.
Recently, there has been increased interest in further developing haptic interfaces, i.e., those that use vibrations, pressure, or temperature as signals.
Haptic information has several advantages: it can be conveyed privately, in an eyes-free manner, and during physical activity~\cite{Roumen2015}.

Despite the increasing interest in haptic information transmission, the modality is far less studied than its visual and auditory counterparts~\cite{Craig1995}. For example, the effects of brief distracting events have been extensively characterized in the visual and auditory domains, yet haptic susceptibility to such distraction remains poorly understood. To make the best use of the haptic modality, such psychological features need to be characterized and then used to inform design.

From a psychological perspective, a sensory event or stimulus is distracting when it captures a person's attention to the detriment of other tasks~\cite{Egeth1997}. For example, a ringing phone may transiently disrupt reading a report. Note that such distraction requires neither perceptual confusion nor masking; basic visual processing is unimpaired by a distracting sound, but attention is redirected away from the visual task to deal with the distracting event or stimulus~\cite{Chun2002}. Impairment can occur even if the person does not look away~\cite{Asplund2010}. Of course, he or she often will.

In addition to intensity, several factors influence whether a stimulus is distracting. These include:

\begin{enumerate}
	\item Expectation. Relatively rare or unpredictable events capture attention~\cite{Egeth1997,Squires1975,Katayama1998,Yamaguchi1991}. Although alerts or notifications are meant to grab attention, there is a cost: The ongoing task is disrupted, leading to slowing or errors~\cite{Chun2002,Asplund2010,Horstmann2015,Vachon2017,Parmentier2011}. In an information theory sense, rare and unpredictable events are surprising~\cite{Mars2008,Horstmann2015}, and their effects are termed "surprise capture". Accordingly, if rare events come to be expected, they become less surprising and therefore less distracting.

	\item Relevance. Events that are relevant to the current task or to broader goals tend to capture attention. For example, if one is waiting for a notification sound, similar sounds will grab one's attention. Such effects are termed "contingent capture", and they have also been shown to disrupt ongoing tasks~\cite{Folk1992,Folk2002}.

	\item Modality. Attention can be captured across modalities (e.g. auditory alerts disrupting visual processing), but distraction tends to be stronger within a given modality (e.g. auditory distraction on an auditory task)~\cite{Ljungberg2012}.

	\item Timing. Distracting effects evolve over time. Indeed, brief distracting events (tens to hundreds of ms) tend to produce powerful yet transient disruptions, which themselves last under a second~\cite{Asplund2010,Horstmann2015,Ljungberg2012}.
\end{enumerate}
These distraction "principles" have been formulated primarily from auditory and visual experimentation, but distracting events may affect haptic tasks in similar ways. Environmental sounds or vibrations (e.g. the "phantom vibrations" of a mobile phone) could be disruptively distracting, as could alerts or notifications that are intended to be helpful~\cite{Egeth1997}. Such alerts can improve performance on a variety of tasks~\cite{Baldwin2012,Merlo2011,Chen2008,Stanley2006,Kochhar2012,Marsalia2016}, but a poorly-timed or overly distracting alert may disrupt processing of the very information to which it was intended to direct attention. Alerts in each sensory domain--including auditory, visual, and haptic--will be experienced more frequently as devices employing them becoming increasingly common.

In this study, we assess the effects of distracting stimuli on two vibrotactile tasks. Each task involves either target detection or pattern discrimination on the fingers, as haptic sensitivity is greatest there~\cite{Cholewiak1984} and smart rings have been developed accordingly. Across four experiments, we explore how these tasks can be disrupted by relatively rare "surprise" distractors, including changes in distraction effects over successive presentations (\textit{Expectation}, see list of factors above). The surprise distractors are not relevant to the primary tasks themselves (\textit{Relevance}), though they have the same modality as the targets (vibrations) in two experiments and have a different modality (sounds) in two others (\textit{Modality}). Finally, we explore the timecourse of distracting effects by varying the time between the surprise distractors and targets (\textit{Timing}).

Our experiments assess the potential importance of attentional capture effects on the detection of vibrotactile information. As such, the study is novel and informative for basic psychology as well as for interactions with haptic devices.

The contribution of this paper is three-fold. Specifically, it provides the following:
\begin{enumerate}
    \item A succinct review of psychological and HCI considerations for distracting events and beneficial alerts. Our focus is on situations with haptic components, though the principles are drawn from and generalize to other sensory modalities.
    \item An empirical investigation of the effects of surprise distraction on vibrotactile task performance. Experiments 1A and 1B show that relatively unexpected vibrotactile and auditory events can impair the detection of a target vibration's location. Experiments 2A and 2B show that these same stimuli harm vibrotactile pattern recognition. The effects of these distraction effects change over time, with strong effects for several hundred milliseconds that then diminish--potentially even becoming beneficial alerts after roughly a second.
    \item A set of design guidelines for avoiding the pitfalls of attentional capture effects on vibrotactile tasks. We also discuss how smart ring systems could be engineered to present information at ideal timings relative to other stimuli.
\end{enumerate}

\section{Related Work}
The current study's context includes prior studies in both HCI and Psychology.
In the former, haptic perception on the fingers, the development of smart rings, distraction caused by auditory or vibrotactile stimuli, and facilatory crossmodal interactions all inform our work.
For the latter, the effects and control of attention, responses to unexpected stimuli (so-called oddball paradigms), and attentional trade-offs that result in detection failures are most relevant to the present study.
Each of these topics is briefly reviewed below.
To the best of our knowledge, no previous study in either HCI or psychology has investigated the effects of distracting stimuli on vibrotactile target detection, yet such effects have great relevance to both fields.

\subsection{Notifications Presented to the Fingers}
Our fingers are highly sensitive to haptic stimulation~\cite{Cholewiak1984}. Spatial acuity (the minimum gap distance between two vibrotactile actuators) is greatest on the fingertip, followed by the remainder of the finger~\cite{Gibson2005}.
As such, smart rings have been developed to convey information to these sensitive areas. A variety of haptic modalities could be used for smart ring notifications; Roumen et al.~\cite{Roumen2015} investigated five of them, finding that vibration was a promising candidate in terms of both reaction time and accuracy. Smart rings also have the potential to convey rather complex, information-rich stimuli. For example, the TactoRing~\cite{Je2017} uses a small tactor to present spatiotemporal patterns, which participants can discriminate well.

\subsection{Distraction from Vibrotactile or Auditory Sources}
To the best of our knowledge, the disruption of haptic tasks has not yet been studied in an HCI setting. Conversely, distraction from haptic (especially vibrotactile) and auditory stimuli has been examined in depth. For example, Zheng et al.~\cite{Zheng2013} examined the attention-capturing effects of haptic actuators. They found settings that would allow these actuators to alert a user with varying degrees of urgency, from gently making aware to interrupting to demanding action. Such considerations have been particularly important for vibrotactile notifications during critical tasks such as driving, where a rare or unexpected vibration intended to inform can be problematically distracting instead~\cite{Marsalia2016}.

Sound can also interfere with haptic perception, both in the lab and applied settings. For example, Qian et al.~\cite{Qian2014} investigated the effect of background noise on complex vibrotactile tacton identification. They compared a controlled environment with fairly constant background noise to an uncontrolled environment with higher unexpected variance in sound levels. In the uncontrolled condition, the haptic recognition rate dropped from 74.3\% to 62.8\%. However, given the background environment's unpredictable and uncontrolled nature, it is not clear whether the effect was triggered by a sudden loud noise (that is, a "surprise") or a general increase in sound. Indeed, auditory noise can interfere with haptic tasks in other settings, as pink noise is frequently used to mask the sound of vibration motors~\cite{Saket2013,Alvina2015}.

\subsection{Facilitatory Interactions Across Modalities}
Haptic stimuli have also been used with the aim of improving performance on critical tasks~\cite{Marsalia2016}. Specifically, Stanley~\cite{Stanley2006} compared unimodal (vibrotactile or auditory only) to bimodal (vibrotactile+auditory) alerts that signaled another vehicle's lane change in a driving simulator. Participants were faster to react to vibrotactile-only stimuli. In another driving simulation study, a vibrotactile warning signalled an otherwise unexpected braking event~\cite{Kochhar2012}. Such a warning decreased reaction times to the event itself.

More generally, the tactile and auditory senses often interact in detection and learning. Hoggan and Brewster proposed audio and tactile crossmodal icons for mobile devices~\cite{Hoggan2007}. Participants learned a set of audio icons (earcons) or tactile icons (tactons). Participants trained in one modality were able to still accurately discriminate the same set of icons in the other modality, suggesting conceptual and sensory connections across the modalities. Similarly, Hoggan et al.~\cite{Hoggan2008} showed that crossmodal congruence was an important determinant of perceived quality of auditory or tactile feedback for visual widgets.

\subsection{Attention's Effects and Control}
Attention is the means by which we manage multiple sensory inputs, selecting and enhancing some while ignoring or suppressing others. Although often under voluntary control, attention can also be directed to a location, time, or item by an environmental stimulus~\cite{Egeth1997}. The stimulus itself may benefit from this involuntary capture of attention, as it is then detected, evaluated, or responded to more quickly or accurately. Alternatively, the environmental stimulus may serve as a cue for another item. Visual, auditory, and haptic cues lead to better response accuracy and speed for subsequently presented targets. These benefits have been found for a variety of lab-based and applied tasks, both within and across modalities~\cite{Carrasco2004,Egeth1997,Baldwin2012,Merlo2011,Chen2008}. Importantly, involuntary cueing effects emerge quickly, within a few hundred milliseconds of cue onset. They usually dissipate quickly as well.

\subsection{Oddball Experiments}
A common approach to studying involuntary attentional control is the oddball paradigm, of which there are visual, auditory, haptic, and crossmodal versions~\cite{Squires1975,Katayama1998,Yamaguchi1991}. In a typical haptic version of the experiement, vibrotactile stimuli (vibrations) are presented serially at a rate of approximately one per second. The majority (80-90\%) of these stimuli are identical standards, while the remaining ones are targets that require a response and/or novel stimuli that should be ignored. The key finding is that these relatively rare stimuli evoke a characteristic electrophysiological response, the P300, which includes a component sensitive to novelty~\cite{Mars2008}. The novelty P300's magnitude is inversely related to stimulus rarity and the degree of expectation, and it rapidly habituates, becoming smaller after repeated presentations~\cite{Yamaguchi1991}. As such, the novelty P300 is thought to index "surprise"--in both the phenominological and information theory senses--and an attentional reorienting response towards it~\cite{Mars2008}. The P300 has also been used in brain-computer interfaces~\cite{Donchin2000}, including those involving rare tactile stimuli~\cite{Brouwer2010}.

\subsection{Attentional Trade-Offs}
Attention is usually studied for its benefits on performance, but attention also has a "dark side": Unattended events frequently go unnoticed~\cite{Chun2002}. For example, detecting a target item in a stream of irrelevant items (~100 ms per stimulus) severely impairs detection of a second target~\cite{Dux2009}. This phenomenon, termed the attentional blink, has been found in tactile~\cite{Hillstrom2002,Dellacqua2006} and cross-modal (e.g., visual and tactile~\cite{Soto2002}) paradigms in addition to its original visual form. More generally, multiple tactile stimuli interfere with one another across both space (fingers) and time due to competition for limited attentional resources~\cite{Craig1995}. Another example is found in the inattentional numbness phenomenon. When one is paying attention to a demanding visual or tactile task, a strong but unexpected vibrotactile stimulus is often not detected~\cite{Murphy2016,Murphy2018}.

Attention's "dark side" can also be evidenced when it is controlled involuntarily. For example, in Surprise-induced Blindness, a rare and unexpected event (akin to an "oddball"; see above) captures attention to the detriment of subsequent target detection~\cite{Asplund2010,Horstmann2015}. These effects are strongest around 300-400 ms, and they dissipate after only a few trials. Auditory attentional capture also induces detection costs~\cite{Vachon2017}, with a comparable timecourse but a much longer-lasting effect across trials. Such distraction also operates across modalities, including the disruption caused by novel vibrotactile stimuli\cite{Parmentier2011,Ljungberg2012}. 

In addition, rare and unexpected stimuli that share a critical feature (e.g. color or vibrational intensity) with the target can also capture attention. Such contingent capture can cause target detection failures as well~\cite{Folk1992,Folk2002}. Given these various attentional capture effects, we designed our present experiments to test for vibrotactile task disruptions caused by relatively rare and unexpected vibrotactile or auditory stimuli.
\section{Experiments 1A and 1B: Effects of Distraction on Vibrotactile Location Detection}
In this first pair of experiments, we sought to understand how a relatively rare and unexpected stimulus (the surprise distractor) would affect the detection of a specific vibrotactile target presented on one of four fingers. We expected that the surprise distractor would have detrimental effects on target detection due to attentional capture~\cite{Parmentier2011,Ljungberg2012,Yamaguchi1991}. These effects could reduce (habituate) with repeated presentations as the surprise distractor became more familiar. Indeed, a surprise distractor could even have facilitatory effects if participants learned that it acts as a temporal cue, predicting target onset~\cite{Parmentier2011,Marsalia2016}.

We designed our task to represent a challenging perceptual scenario. Target detection was made difficult through the choice of stimuli, which were short pulses on different fingers without an intervening pause~\cite{Craig1995, Hillstrom2002}. In Experiment 1A, we used a brief, wrist-based vibration as the Surprise distractor. In Experiment 1B, we used a sound. Otherwise, the experiments contained virtually identical apparati, procedures, tasks, and stimuli.

\subsection{Common Apparatus}
\begin{figure}[!]
	\centering
  \includegraphics[width=.80\columnwidth]{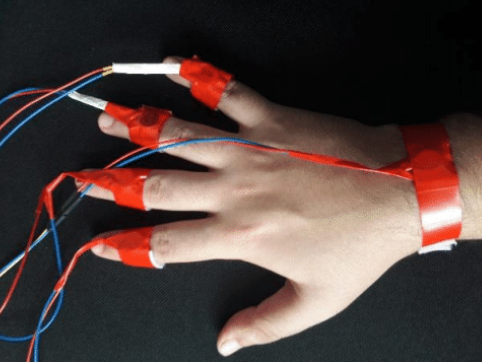}
  \caption{%
    Apparatus used in Experiment 1A, with four motors on the fingers and one on the wrist. In Experiment 1B, the wrist motor was not attached.
    \label{fig:1a-apparatus}%
  }
\end{figure}

An adjustable ring-like device was attached to the middle phalanx of each finger of the dominant hand (excluding the thumb). Using the middle phalanx allowed for more spatial separation between the devices, and the middle and proximal phalanges have similar sensitivity~\cite{Weinstein1962}. (Nevertheless, note that we used the proximal phalanx, a more natural location for smart ring placement, in Experiment 2.) Each device contained a small, coin-type vibration motor. In Experiment 1A, an additional motor was attached to the wrist (Figure~\ref{fig:1a-apparatus}). All motors were attached to the same hand because same-hand interference is better characterized in the tactile literature~\cite{Craig1995,Hillstrom2002}, and we sought to test for potential interference.
Each vibration motor had a typical operating amplitude of 1.3 G (Precision Microdrives 310-103; data sheet available at \href{https://www.precisionmicrodrives.com/product/datasheet/310-103-10mm-vibration-motor-3mm-type-datasheet.pdf}{https://www.precisionmicrodrives.com/product/datasheet/310-103-10mm-vibration-motor-3mm-type-datasheet.pdf}). The motors were controlled by an Arduino Uno microcontroller. As the Arduino provided 5 V with a maximum current of 40 mA, the motors were driven outside their rated typical values of 3 V and 58 mA. We therefore tested the motors with our setup to record their frequencies and relative amplitudes for each experiment (see the Common Task and Stimuli sections for both Experiment 1 and Experiment 2).

The Arduino Uno itself was controlled by a 2013 iMac running our Python-based experimental software. Participant responses were recorded through a standard USB keyboard. Participants rested the fingers of their dominant hand on four response keys (index finger on F, middle finger on G, ring finger on H, and small finger on J; reversed for left handers). This setup meant that participants indicated the target vibration location by using the stimulated finger itself, thereby reducing confusion; the raised bumps on the F and J keys helped participants keep their hands on the appropriate set of keys (see additional notes on this design choice in the Experiment 1B introduction below). The response hand was covered by a box so that motor vibration was not visible. Participants also wore Sony MDR-ZX110 over-ear headphones connected to the computer's 3.5mm headphone jack. Participants were told that the purpose of the headphones was to block the noise of the vibrating motors, which was one of their primary functions. No sound was played through the headphones, save the auditory surprise distractors in Experiment 1B.

\subsection{Common Procedure}
Before the experiment began, participants were briefed on the task and provided informed consent. The consent procedure and protocols for all four reported experiments were approved by the local IRB.
After wearing the apparatus, each participant was trained in its use. The experimenter generated a stimulus and then asked the participant to identify on which finger the target was presented.
The experiment was divided into 7 blocks, each with 24 trials.
The first block was for practice and contained no surprise stimuli.
In each of the 6 subsequent blocks, 4 trials (16.67\% of the total) contained an unexpected alert (surprise stimulus).
The rarity of the surprise, as well as its timing relative to the target, was chosen based on previous literature~\cite{Asplund2010,Horstmann2015,Folk2002,Vachon2017,Yamaguchi1991}.
Participants were encouraged to take breaks between blocks and completed the experiment in approximately 30 minutes. Participants were compensated with either payment (3.7 USD) or optional research participation credit for a psychology course.

\subsection{Common Task and Stimuli}
Participants rested the fingers of their dominant hand on the four response keys. They then initiated each trial by pressing the keyboard's space key. A sequence of eight 200 ms vibrational pulses followed. The location (finger) of each pulse was random, with the restriction that a finger would not receive two sequential pulses.
One of the eight pulses (the target, item 3-7 in the sequence) was driven with a pulse width modulation (PWM) value of 100\%, whereas the other pulses were driven at PWM 40\%. Stimulus measurements confirmed that the relative vibration amplitudes were roughly as intended, with 40\% PWM stimulus amplitude approximately 35\% of the 100\% PWM stimuli. The 100\% PWM stimuli (targets and surprise distractors) vibrated at approximately 130 Hz, a frequency at which the motor would produce minimal bone conductance or direct auditory perception~\cite{Mcbride2017}. The 40\% PWM stimuli vibrated at approximately 40 Hz. Although bone conductance is more likely at this frequency, neither the target's location nor occurence would be revealed by such stimuli. We selected the PWM settings and vibration durations following pilot sessions, choosing values that provided a challenging yet feasible task. Participants indicated the location (finger) of each trial's target vibration using the stimulated finger itself.
When present, the surprise distractor began 350 milliseconds before the target vibration. At least two trials without a surprise distractor (target-only trials) occurred between any two trials with a surprise distractor (labelled as "surprise trials").

\subsection{Common Design}
A $4 \times 2$ within-subject design was used with three independent variables, including a continuous one: \textit{Trial Type \{Target Only, Surprise\}}, \textit{Finger \{ Index, Middle, Ring, Pinky\}}, and \textit{Session Time \{trial position within the session\}}. For visualization puposes, we binned Session Time into three periods (Early: Blocks 2 and 3, Middle: Blocks 4 and 5, Late: Blocks 6 and 7). Trial Type and Finger were fully crossed and randomized within each Block. As such, each participant experienced 4 fingers $\times$ [1 (training block) + 6 (test blocks)] $\times$ 6 repetitions = 168 trials, including 24 trials with a surprise distractor.
Our dependent variable was accuracy. A trial's response was considered successful if the participant was able to correctly identify on which finger the target vibration occurred.

\subsection{Common Analytical Approach}
Data preparation, statistical analysis, and visualizations were all implemented in RStudio version 1.0.136 (R Foundation for Statistical Computing) running R version 3.2.4. For descriptive statistics, we report marginal means and standard deviations or standard errors of the mean. Before calculating descriptive and inferential statistics for the sample, we removed participants whose detection performance in target-only trials was not reliably above chance performance, a procedure used in similar studies~\cite{Murphy2016,Murphy2018,Obana2019}. We reasoned that we could not assess the effects of surprise distractors on performance from such individuals. Data from participants whose performance was below 40\% correct detection on any finger were removed from the sample. Given that targets were delivered to each finger in 36 target-only trials per participant, the 40\% cutoff identifies performance significantly above chance (25\%) based on proportion tests. Furthermore, we examined the pattern of errant responses to ensure that low participant performance was not due to swapped motor positions (experimenter error) or consistent perceptual or response biases (e.g. consistently reporting index finger stimulation as middle finger stimulation). We found only one such case, which is described in Experiment 1B.

Given each trial's binary detection outcome, we constructed generalized linear mixed-effect models for our primary analyses. The factors were Trial Type (2 levels; surprise trial or target-only trial), Session Time (continuous), and Finger (4 levels, one per finger location of the target). By-subject random slopes and intercepts for the categorical factors (Trial Type and Finger) were included in the error terms~\cite{Barr2013}. The resulting logistic regression models were fit using glmer() in the lme4 package (version 1.1-17)~\cite{Bates2015}. They were then assessed using the Anova() function in the car package (version 3.0-2), which generates tables of Type II Wald chi-squared tests~\cite{Fox2011}.
Follow-up pairwise tests were conducted on the estimated marginal means using emmeans() in the emmeans package (version 1.3.1). In addition, we conducted a secondary analysis on the types of errors made, either those to an adjacent finger (near) or another finger (far). As the dependent measure for this analysis was response rate (continuous), we constructed linear mixed-effect models with factors of Trial Type and Error Distance (near or far). The models were fit using lmer() in the lme4 package. To control for multiple comparisons, false discovery rate (FDR) correction was applied to all p-values in the study~\cite{Benjamini1995}.

\section{Experiment 1A: Effects of Vibrotactile Surprise Distractors on Target Detection}
In the first experiment, the relatively rare and unexpected stimulus (the surprise distractor) was a single 200 millisecond vibration (100\% PWM) presented on the back of the wrist, mimicing the placement of a smartwatch.
The surprise distractor had the same duration as all other vibration pulses, as well as the same intensity as the target vibration.
The apparatus, procedure, task, stimuli, and design were described above.

\subsection{Participants}
Forty-two participants (27 female, 39 right-handed), aged 18-32 ($M=22.6$) took part in the experiment.
Owing to apparatus failures or experimenter error, the data from 10 participants could not be used. In addition, eight participants were excluded for failing to achieve at least 40\% target detection for each finger.
The final sample included 24 participants $\times$ 168 trials = 4,032 trials, including 576 with a surprise distractor.

\subsection{Results}
\begin{figure}[!]
	\centering
  \includegraphics[width=.50\columnwidth]{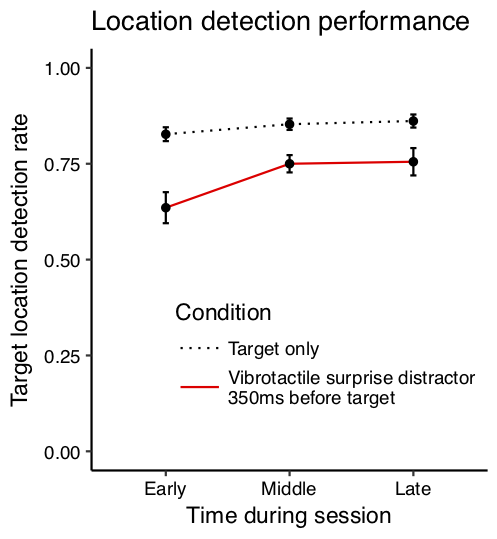}
  \caption{%
    Recognition rate for Experiment 1A. Error bars represent the standard error of the mean. Each value for time during session represents two blocks, the training block is not counted.
    \label{fig:1a-results}%
  }
\end{figure}

After the data from Block 1 (practice) had been discarded, we visualized the results in Figure~\ref{fig:1a-results}.

\paragraph{Trial Type.}
Although target detection performance was high, unexpected wrist vibrations substantially impaired performance (main effect of Trial Type: $\chi^{2}(1)=59.8, p<.001$). Participants successfully detected the target more often without the Surprise distractor ($M=84.7\%$) than with it ($M=71.4\%$).

\paragraph{Session Time.}
Performance also improved over time (main effect of Session Time: $\chi^{2}(1)=11.9, p=.001$).
Performance on target-only trials rose from 82.7\% during the early part of the experiment to 86.1\% during the late part, whereas accuracy rose from 63.5\% to 75.5\% for surprise trials.

\begin{figure}[!]
	\centering
  \includegraphics[width=.80\columnwidth]{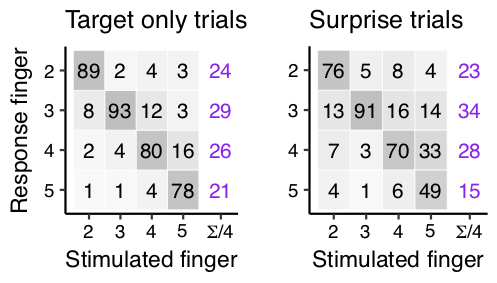}
  \caption{%
    Confusion matrices for Experiment 1A. Stimuli are columns, responses rows. Accuracy is in \%.
		The $\frac{\sum}{4}$ column shows how frequently each response was chosen. Equally distributed answers have a value of 25.
		Higher values suggests that this answer is chosen more often. Finger 2 is index finger, 3 is middle finger, 4 is ring, 5 is pinky.
    \label{fig:1a-confusion}%
  }
\end{figure}

\paragraph{Finger.}
We found a significant main effect of Finger on accuracy ($\chi^{2}(3)=25.6, p<.001$). To better understand this effect, we constructed confusion matrices (Figure~\ref{fig:1a-confusion}), which revealed participants' error types. Most errors involved spatial confusion, in which the finger adjacent to the target location was selected ("near" errors). Note that if errors were random, one would expect an approximately equal frequency of "near" and "far" errors; six cells of the error matrix contribute to each error type. A follow-up linear mixed-effects model (see above) revealed significant main effects of both Trial Type ($\chi^{2}(1)=42.8, p<.001$) and Error Distance ($\chi^{2}(1)=24.9, p<.001$), with no significant interaction ($\chi^{2}(1)=0.002, p=.98$). The presence of a surprise distractor increased both types of errors by approximately the same amount. Near errors increased from 11.7\% to 19.3\%, whereas far errors increased from 3.5\% to 9.4\%. These results indicate that the surprise distractors primarily induced more random guessing, not increased spatial confusion. We suggest that during most surprise trials, the target was missed completely, not merely degraded~\cite{Asplund2014,Sergent2004}.

\paragraph{Interactions.}
We found a significant Trial Type $\times$ Finger interaction ($\chi^{2}(1)=12.8, p=.010$). During target-only trials, participants were more accurate with their index ($M=88.8\%$) and middle finger ($M=92.6\%$) as compared to their ring ($M=79.9\%$) and pinky ($M=77.6\%$). Post-hoc pairwise tests confirmed these performance scores were significantly different from one another ($p<.003$), save ring versus pinky ($p=.28$). Performance was also significantly different across finger pairs during surprise trials (all $p<.01$), save index versus ring ($p=.87$). Detection problems were particularly acute on the pinky ($M=48.6\%$), with 33.3\% of targets delivered to it reported as ring finger stimulation. Finally, significant impairments due to the surprise distractors were found on all fingers ($p<.035$), with the exception of the middle one ($p=.49$). The middle finger received more responses overall during surprise trials, so response bias could account for the null effect.
No other interactions were significant (Trial Type $\times$ Session Time: $\chi^{2}(1)=2.45, p=.17$; Session Time $\times$ Finger: $\chi^{2}(3)=5.59, p=.19$; Trial Type $\times$ Session Time $\times$ Finger: $\chi^{2}(3)=5.90, p=.17$).

\subsection{Discussion}
Our results suggest that unexpected vibrotactile stimuli have an impact on user's ability to perform the task, with an average decrease in accuracy of 13.3 percentage points. Importantly, this deficit is fairly consistent over time, with a 10.6 points effect still present late in the session. Moreover, despite repeated exposure to the surprise distractor and its consistent prediction of the target's arrival, participants did not use the alert to improve their performance.
Our experiment also highlights perception differences between the fingers, as we observed an effect of Finger on accuracy. Performance on the index and middle finger was consistently better, even during surprise stimulus presentation. In contrast, target detection on the pinky began lower during target-only trials (77.6\%) and was further reduced during surprise trials (48.6\%). Although the overall performance differences across the fingers are clear, the additional deficits caused by surprise distraction are more difficult to interpret. We consider these ideas further in the General Discussion.

\section{Experiment 1B: Effects of Auditory Surprise Distractors on Target Detection}
In Experiment 1A, we showed that a relatively rare and unexpected vibrotactile stimulus could disrupt target detection, even though the surprise distractor was separately both spatially and temporally from the target. The shared modality (as well as shared features) of the target and surprise distractor may have rendered participants unable to ignore the surprise distractor. To investigate whether disruption could occur even across modalities, we investigated the effects of unexpected auditory distractors on vibrotactile target detection. Such distraction is ecologically relevant, as we are bombarded with both relevant and irrelevant noises in our daily life, ranging from conversations to chirping birds to device notifications.

For Experiment 1B, we used a set of such sounds that have been shown to cause distraction to the detriment of auditory task performance~\cite{Obana2019}. These sounds included spoken letters, environmental noises, and synthetically produced pitches. Each sound was used as a surprise distractor in a given surprise trial, with the sound presented through headphones 350 milliseconds before the target vibration. For each surprise trial, the surprise distractor was randomly drawn from the set of 80 sounds. Changing the surprise distractor has been found to increase the longevity of the auditory effects across successive surprise trials (Surprise-induced Deafness), with few apparent differences in the effects across the various sound types~\cite{Obana2019}. All sounds were compressed to 110 milliseconds and normalized to ensure a comparable energy profile~\cite{Font2013}. Stimulus measurements confirmed the roughly equal volumes, which ranged between 60 and 70 dB~\cite{Shen2006,Obana2019}. The apparatus, procedure, task, stimuli, and design were otherwise as described above.

Second, the first six participants in the sample responded with their non-dominant hand, while their dominant hand received the vibrotactile stimulation. Although instructed to respond with the corresponding finger (e.g. left index for right index stimulation), we found that 3 of 6 participants had consistently reversed their mappings (based both on data analysis and participant reports). These individuals had instead responded with corresponding spatial positions (e.g. left pinky, the rightmost finger of the left hand, for right index stimulation). To ensure more consistent performace with reduced input errors, we adjusted the procedures so that the dominant hand was used for both stimulation and response. As baseline task performance (after correction, when applicable) for these first six participants was qualitatively similar to the remainder of the sample, we included them in the final sample. Their inclusion or exclusion did not affect the final results or our conclusions. Finally, as Experiment 1A's data collection actually started slightly after Experiment 1B's began, we could use the improved procedure throughout.

\subsection{Participants}
Twenty-seven participants (16 female, 25 right-handed), aged 18-29 ($M=21.5$) took part in this experiment. The finger-bands for one left-handed participant were errantly reversed; the data from this participant were included after correction. Seven participants were excluded from the final sample for failing to achieve at least 40\% target detection on one or more fingers, performance that was due neither to errant motor placement nor consistent perceptual biases.
The final sample included 20 partipants $\times$ 168 trials = 3,360 trials, including 480 with a surprise stimulus.

\subsection{Results}
\begin{figure}[!]
	\centering
  \includegraphics[width=.50\columnwidth]{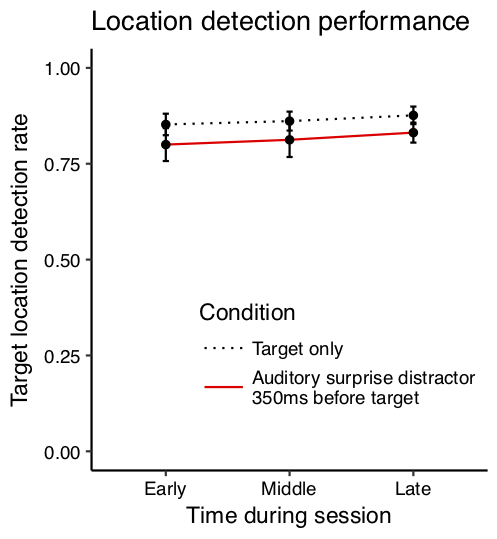}
  \caption{%
    Recognition rate for Experiment 1B. Error bars represent the standard error of the mean. Each value for time during the session represents two blocks.
    \label{fig:1b-results}%
  }
\end{figure}

After the data from Block 1 (practice) had been discarded, we visualized the results in Figure~\ref{fig:1b-results}.

\paragraph{Trial Type.}
Target detection performance during target-only trials was high ($M=86.3\%$), and surprise distractors impaired performance ($M=81.5\%$). The impairment was numerically slight, albeit still significant (main effect of Trial Type: $\chi^{2}(1)=9.51, p=.004$).

\paragraph{Session Time.}
Performance did not significantly vary across the session (main effect of Session Time: $\chi^{2}(1)=1.10, p=.37$).

\begin{figure}[!]
	\centering
  \includegraphics[width=.80\columnwidth]{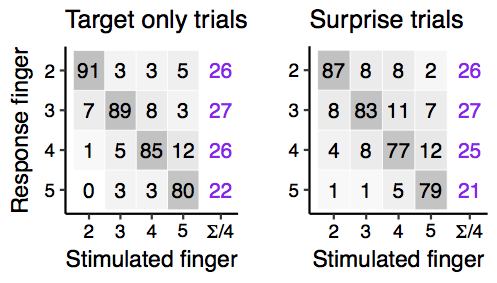}
  \caption{%
    Confusion matrices for Experiment 1B. Stimuli are columns, responses rows. Accuracy is reported as a percentage. Finger 2 is index finger, 3 is middle finger, 4 is ring, 5 is pinky.
    \label{fig:1b-confusion}%
  }
\end{figure}

\paragraph{Finger.}
Performance did not significantly vary across fingers (main effect of Finger: $\chi^{2}(3)=6.25, p=.16$).
Averaged across all trials (target-only and surprise), performance ranged from 90.5\% for the index to 79.9\% for the pinky. Individual finger performance is reported in Figure~\ref{fig:1b-confusion}. Consistent with Expt. 1A, most errors involved spatial confusion ("near" errors). A follow-up linear mixed-effects model (see above) revealed significant main effects of both Trial Type ($\chi^{2}(1)=7.99, p=.009$) and Error Distance ($\chi^{2}(1)=5.99, p=.026$), with no significant interaction ($\chi^{2}(1)=0.00, p=.99$). As before, the presence of a surprise distractor increased both types of errors. Near errors increased from 9.8\% to 13.1\%, whereas far errors increased from 3.9\% to 5.4\%.

\paragraph{Interactions.}
We did not find any significant interactions in this experiment (all $p>.23$).

\subsection{Discussion}
\paragraph{Experiment 1B}
Our auditory surprise distractors impaired performance, with a small but statistically significant loss of accuracy (4.9 percentage points) that appeared to be stable across time.
Performance differences across individual fingers showed a pattern similar to Experiment 1A, though neither these differences nor the interaction with Trial Type were significant. Although these null results could reflect a true absence of such effects, they could represent a lack of power for detecting more small-to-moderate differences. We consider this issue further in the General Discussion.

Experiment 1B contained two other limitations worth noting, and we addressed each in the design for Experiment 2B. First, a different surprise distractor was used for each trial, which may have contributed to the persistent target detection deficit. Second, in subsequent testing for a different study with a similar apparatus, we found that participants reported that a vibrotactile and an auditory stimulus began simultaneously when the software command to the vibration motor was issued approximately 33 ms before its command to play a pure tone sound. This small timing error likely reflects both different hardware latencies and perceptual factors. Regardless, we still observed a target detection deficit, though its magnitude relative to Experiment 1A's may reflect differences other than modality alone.

Despite these limitations, the primary conclusion of the experiment is clear: Unexpected sounds can disrupt vibrotactile target detection. To avoid the deleterious effects of auditory surprise distraction, one could design vibrotactile systems that also detect sounds. The pattern's presentation could then be delayed in the presence of a potentially distracting sound; this strategy would increase accuracy at the cost of time. We further discuss this possibilty for system design in the General Discussion.

\paragraph{Experiments 1A and 1B}
Our first pair of experiments shows that a surprise distractor, a relatively rare and unexpected stimulus, can impair participants' ability to identify a subsequent target.
This effect was found for both vibrotactile and auditory surprise distractors. Although the vibrotactile surprise distractor was separated both spatially and temporally from the target, their shared modality likely contributed to the larger surprise-related impairment~\cite{Hillstrom2002,Soto2002,Dellacqua2006}.
Participants may also have unwittingly shifted their attentional focus to the wrist because the surprise's features (a relatively strong vibration) matched the target's, thereby leading to contingent attentional capture~\cite{Folk1992,Folk2002}.
Finally, the effect's persistence throughout each session suggests that participants do not--and perhaps cannot--ignore the surprises or use them as predictive cues of the target's imminent arrival.

Our experiments also show that vibrotactile target detection performance may differ across fingers.
Previous work suggests that spatial acuity and force detection is better on the index and middle fingers~\cite{VegaBermudez2001,King2010,Sathian1996}, which is consistent with the findings reported here.
In addition, surprise distractors may affect detection across the fingers differently, with large effects on the pinky and more moderate effects on the index and middle fingers. Such differences, however, may simply reflect baseline performance differences across the fingers.
Regardless, we suggest using the index and middle fingers for smart ring notification delivery.

These initial experiments also contained limitations, which we addressed in Experiments 2A and 2B. First, although pilot tests indicated that the headphones effectively blocked the sounds from the vibration motors, it is possible that they did not completely block such sounds for all participants. That said, it is unlikely that these sounds were the primary cause of the observed deficits. Unexpected vibrotactile distractors (Experiment 1A) caused worse target detection performance as compared to auditory distractors (Experiment 1B), despite the latter's greater volume. In Experiment 2, white noise was presented through the headphones. Second, in Experiment 1, the durations of the surprise distractors differed across the modalities, and auditory surprise distractors varied while only one vibrotactile surprise distractor was used. These differences were removed in Experiment 2, yet it is still challenging to draw strong conclusions about the relative magnitudes (and importance) of vibrotactile and auditory surprise distraction on target detection. Nevertheless, the presence of both types of deficits is important, as is their relative magnitude for addressing potential confounds (e.g. noise from the vibration motors). We further discuss comparisons across modalities in the Limitations section following the General Discussions.

In addition to addressing these limitations, the aim of Experiment 2 was to use a more HCI-relevant experimental design to explore the timing between the surprise distractor and its target. Characterizing this timing relationship is important for understanding the surprise distraction effects, as well as for designing systems that mitigate problematic attentional capture deficits, perhaps while also using alerting notifications effectively.
\section{Experiments 2A and 2B: Effects of Surprise Distractors on Vibrotactile Pattern Recognition}
In this second pair of experiments, we investigated how different timings between the surprise distractor and target affect vibrotactile pattern discrimination. To increase our study's relevance to real-world device design, we also used a different apparatus and set of target vibrations inspired by relevant HCI work~\cite{Kaaresoja2005,Alvina2015,Saket2013,Yatani2009,Yatani2012}.
The apparatus was a single motor worn on the index finger, positioned similarly to a smart ring on the proximal phalanx~\cite{Roumen2015,Je2017}.
The task itself required participants to identify a target pattern (1 of 4). It therefore mimicked an actual notification system.
Experiments 2A (vibrotactile surprise distractor) and 2B (auditory surprise distractor) shared many procedural, task, design, and analysis components, which we describe below.

\begin{figure}[!]
	\centering
  \includegraphics[width=.80\columnwidth]{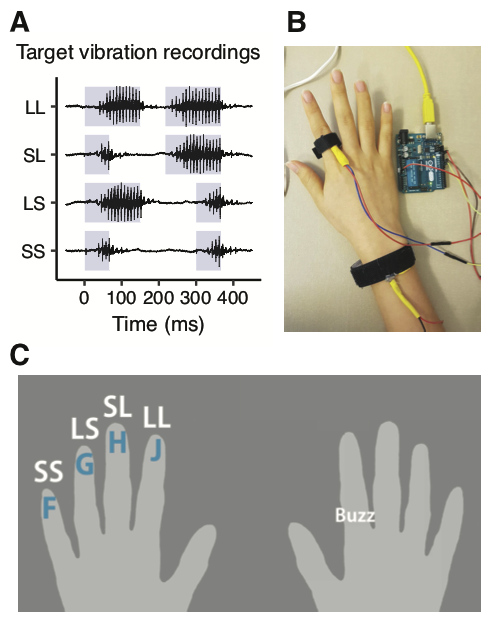}
  \caption{%
    (A) Vibration patterns used in Experiment 2. The waveforms were recorded with a microphone placed near the suspended vibration motor. Areas shaded in light blue represent the time periods when power was supplied to the vibration motor.
    (B) Apparatus used in Experiment 2. In Experiment 2B, the wrist motor was not attached.
    (C) Participants made responses after discriminating 4-alternative forced choices (LL, SL, LS, or SS) using 'j', 'h', 'g', or 'f' keys respectively.
    Illustration for a right-handed participants inputting answers with their non-dominant hand.
    \label{fig:2a-apparatus}%
  }
\end{figure}

\subsection{Common Apparatus}
A 5V vibration motor in an adjustable ring-like device, similar to those used in Experiment 1, was attached to the proximal phalanx of the index finger of each participant's dominant hand.
In Experiment 2A, an additional motor was attached to the wrist (Figure~\ref{fig:2a-apparatus}-B).
Experiment 1's 2013 iMac, keyboard, headphones, and Arduino Uno microcontroller (driving the motors) were used in Experiment 2. The experiment itself was coded in PsychoPy2 v1.83.03~\cite{Peirce2007}.

\subsection{Common Procedure}
Before the experiment began, participants were briefed on the task and provided informed consent.
They then wore the apparatus and familiarized themselves with the four vibration patterns.
Subsequently, they were trained on sequences of 8 target-only trials, which were repeated until accuracy reached 75\%.
After training, participants completed 2 blocks of 84 trials each. Each block contained 12 surprise trials (14.29\% of the total), with 4 surprises per SOA (stimulus onset asynchrony; SOA is the time between the onset of the surprise distractor and the onset of the target).
During each trial, white noise was played through the headphones (64 dB) as an additional control for any sound generated by the motors.
Participants completed the experiment in approximately 30 minutes. They were compensated with either payment (3.7 USD) or optional research participation credit for a psychology course.

\subsection{Common Task and Stimuli}
Participants initiated each trial by pressing the keyboard's space key.
A white fixation cross was then presented in the center of the screen for a variable time beteween 1500 and 3000 milliseconds.
The target vibration pattern then followed. Each 366 millisecond target pattern consisted of two pulses (PWM value of 100\%), each either short (66 milliseconds) or long (150 milliseconds). Stimulus measurements showed that each stimulus had a vibration frequency of approximately 110-130 Hz, and their durations were slightly shorter than the Arduino's driving signals. Driving signals and stimulus recordings for the four combinations (long-long, long-short, short-long, and short-short) are shown in Figure~\ref{fig:2a-apparatus}-A.
Participants were next prompted to choose the response key that corresponded to the target pattern; the prompt was a diagram of these mappings (see Figure~\ref{fig:2a-apparatus}-C).
During surprise trials, the suprise distractor began either 117, 350, or 1050 milliseconds before the target vibration. These onset delays were referred to as SOA (stimulus onset asynchrony) values. At least two trials without a surprise distractor occurred between any two surprise trials.

\subsection{Common Design}
A $4 \times 4$ within-subject design was used with three independent variables, including a continuous one: \textit{Trial Type \{ No Surprise, 117 ms SOA, 350 ms SOA, 1050 ms SOA \}}, \textit{Pattern \{ Long-Long (LL), Long-Short (LS), Short-Long (SL), Short-Short (SS) \}}, and \textit{Session Time \{trial position within the session\}}. For visualization puposes, we binned Session Time into three periods (Early: first 56 trials, Middle: next 56 trials, Late: last 56 trials), comparable to the binning for Experiment 1. Trial Type and Pattern were fully crossed and randomized within each Block. As such, each participant experienced 4 fingers $\times$ 2 blocks $\times$ 21 repetitions = 168 trials, including 24 trials with a surprise distractor, with 8 per SOA.
Our dependent variable was accuracy, i.e. correct identification of the target vibration pattern.

\subsection{Common Analytical Approach}
Similar to Experiment 1, we first excluded data from participants who failed to reach 40/
Follow-up pairwise tests were conducted on the estimated marginal means using emmeans() in the emmeans package (version 1.3.1). In addition, the emtrends() function was used to assess differences in slopes for the continuous Session Time factor.

\section{Experiment 2A: Effects of Vibrotactile Surprise Distractors on Target Pattern Recognition}
In this experiment, the surprise distractor was a 117 ms vibrational pulse (100\% PWM) of the wrist motor. The apparatus, procedure, task, stimuli, and design were described above.

\subsection{Participants}
Forty-three participants (21 female, 39 right-handed), aged 18-27 ($M=21.7$) took part in the experiment. Data from 4 participants were excluded from the final sample due to procedural errors, whereas data from 3 participants were excluded because performance was not at least 40\% for each pattern during target-only trials.
The final sample thus included 36 participants $\times$ 168 trials = 6,048 trials, including 864 with a surprise distractor.

\subsection{Results}
\begin{figure}[!]
	\centering
  \includegraphics[width=.50\columnwidth]{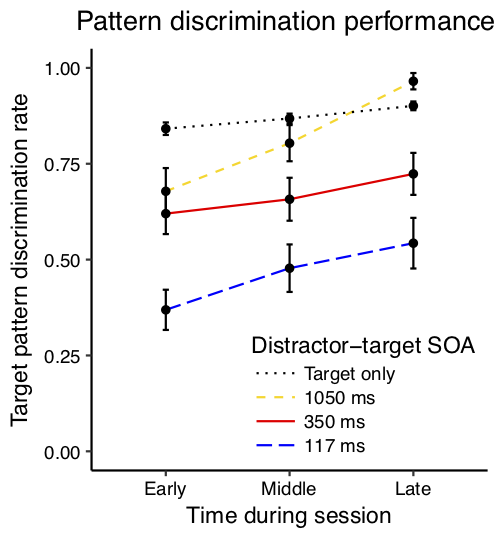}
  \caption{%
    Recognition rate for Experiment 2A for No Surprise (target-only) trials (black dotted line), and for Surprise trials at each SOA value. Error bars represent the standard error of the mean.
    \label{fig:2a-results}%
  }
\end{figure}

The results are summarized in Figure~\ref{fig:2a-results}.

\paragraph{Trial Type.}
There was a significant main effect of Trial Type ($\chi^{2}(3)=241, p<.001$). Indeed, the surprise distractor effects were considerable: Compared to the high baseline performance in No Surprise (target-only) trials ($M=87.0\%$), performance in the 117 ms SOA surprise trials was 40.8 percentage points lower ($M=46.2\%$), and performance in the 350 ms SOA surprise trials was 21.0 percentage points lower ($M=66.0\%$). Post-hoc tests showed that each of these conditions was significantly different from the other, and the three different SOA conditions were also significantly different from one another (all $p<.001$). In contrast, the 1050 ms SOA surprise trials had performance that was not significantly different from the No Surprise trials ($p=.28$).

\paragraph{Session Time.}
We also found a significant main effect of Session Time ($\chi^{2}(1)=54.6, p<.001$) on accuracy. Performance gradually increased across the session.

\begin{figure}[!]
	\centering
  \includegraphics[width=.80\columnwidth]{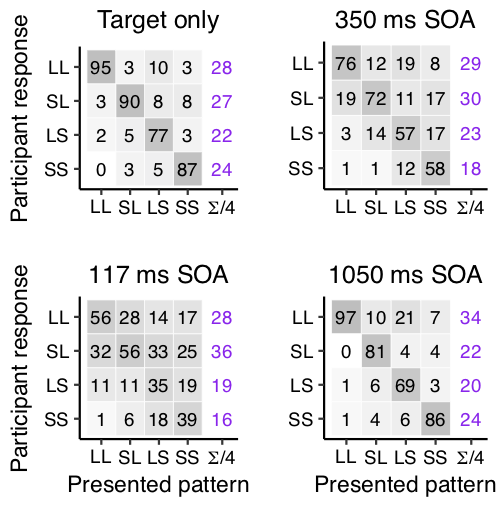}
  \caption{%
    Confusion matrices for Experiment 2A. Stimuli are columns, responses rows. Accuracy is in \%.
		The $\frac{\sum}{4}$ column shows how frequently each response was chosen. Equally distributed answers have a value of 25.
		Higher values suggests that this answer is chosen more often. Finger 2 is index finger, 3 is middle finger, 4 is ring, 5 is pinky.
    \label{fig:2a-confusion}%
  }
\end{figure}

\paragraph{Pattern.}
There was a significant main effect of Pattern ($\chi^{2}(3)=47.4, p<.001$). Post-hoc comparisons were all significant ($p<.002$) except SL vs. SS ($p=.19$). Individual pattern performance is reported in Figure~\ref{fig:2a-confusion}.

\paragraph{Interactions.}
We found a significant interaction of Trial Type $\times$ Session Time ($\chi^{2}(3)=17.6, p=.001$). Pairwise tests of each Session Time slope by Trial Type condition revealed a difference between the 117 ms SOA and No Surprise conditions that did not reach statistical significance ($p=.083$) and a difference between the 1050 ms SOA and No Surprise conditions that was statistically significant ($p<.001$). Performance improved faster for the surprise trials than the target-only ones. For the 1050 ms SOA condition, the initial deficit actually became a significant 6.4 percentage points benefit by the final third of the session (Wilcoxon signed ranks tests on effect in early and late bins, $p<.05$).
Finally, the interaction of Trial Type $\times$ Pattern did not reach statistical significance ($\chi^{2}(9)=16.6, p=.086$). As Figure~\ref{fig:2a-confusion} shows, the surprise distractor effects were grossly consistent across the vibration pattern targets.

\subsection{Discussion}
Foremost, the results show that surprise distractors greatly affect performance, but the timing of these surprise distractors is crucial to their effects.
For the two shortest SOA values, accuracy was a considerable 40.8 percentage points (117 ms SOA) or 21.0 percentage points (350 ms SOA) lower than the target-only (No Surprise) trials baseline.
In contrast, the surprise effects for the 1050 ms SOA were no different from baseline, and these effects changed from a significant deficit to a significant benefit across the session. Note that this increase in performance was on top of a general improvement in task accuracy across the session.

Examination of the confusion matrices revealed additional findings of note. Foremost, participants tended to report the second pulse as long. Consequently, discrimination was higher for the LL and SL patterns regardless of condition. The pattern of errors (misidentifications) reflected the same tendency: Errant reports of a long second pulse were about three times as frequent as errant reports of a short second pulse for each condition (target-only: 2.9, 117 ms: 3.1, 350 ms 2.9, 1050 ms: 3.0). This bias may represent a mixture of perceptual and mechanical effects. Regardless, target-only discrimination performance was reasonably high for all patterns, so participants could perform the task well.

The similar pattern of errors across conditions also sheds light on how the surprise distractors impair target discrimination. One possibility is that the the surprise distractors interrupt target processing, perhaps by drawing attention away from the target. A second possibility is that each surprise distractor becomes integrated into the target pattern, especially at the short 117 ms SOA. On this account, the perceived length of the target's first vibration would be increased, thereby causing more LL and SL reports. We did not observe such an increase in these reports (see Figure~\ref{fig:2a-confusion}). Indeed, in the 117 ms SOA condition, the SL pattern was mistaken for LL at almost the same rate as the reverse. Furthermore, both LS and SS were often errantly reported as SL, supporting the interruption account instead of integration. The distinct spatial locations of the surprise distractor and target patterns may have prevented integration.

\section{Experiment 2B: Effects of Auditory Surprise Distractors on Target Pattern Recognition}
In Experiment 2B, we tested the effects of an auditory surprise distractor on pattern recognition. The surprise distractor was a 110 millisecond gliding sound played through headphones (measured at 64 dB). In psychophysical testing for a different study, we found that participants reported a vibrotactile stimulus and an auditory stimulus to have begun simultaneously when the software command for the former commenced 33 ms earlier. We therefore included this offset in the experiment code, and we confirmed that it produced the desired SOAs using stimulus recordings.
The apparatus, procedure, task, stimuli, and design are otherwise described above.

\subsection{Participants}
Thirty-one participants (12 female, 31 right-handed), aged 18-26 ($M=21.3$) took part in the experiment. The data from 6 participants was excluded because performance was not at least 40\% for each pattern during target-only trials.
The final sample thus included 25 participants $\times$ 168 trials = 4,200 trials, including 600 with a surprise distractor.

\subsection{Results}
\begin{figure}[!]
	\centering
  \includegraphics[width=.50\columnwidth]{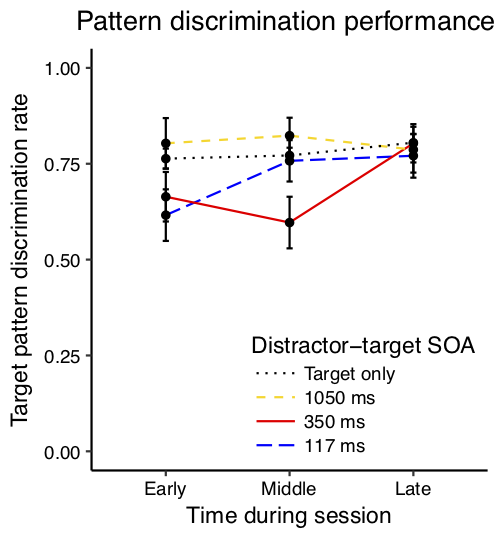}
  \caption{%
    Recognition rate for Experiment 2B for No Surprise (target-only) trials (black dotted line), and for Surprise trials at each SOA value. Error bars represent the standard error of the mean.
    \label{fig:2b-results}%
  }
\end{figure}

The results are summarized in Figure~\ref{fig:2b-results}.

\paragraph{Trial Type.}
There was a significant main effect of Trial Type ($\chi^{2}(3)=10.6, p=.026$). Performance was only moderately affected by the surprise distractor, with a 9.0 percentage point deficit in the 350 ms SOA condition ($M=69.0\%$) relative to the target-only baseline ($M=78.0\%$). Post-hoc tests revealed this difference to be significant ($p=.034$), but performance for neither the 117 ms SOA condition ($M=72.5\%$) nor the 1050 ms SOA condition ($M=80.0\%$) was significantly different from baseline (both $p>.17$)

\paragraph{Session Time.}
We also found a significant main effect of Session Time ($\chi^{2}(1)=14.5, p<.001$) on accuracy. Performance gradually increased across the session.

\begin{figure}[!]
	\centering
  \includegraphics[width=.80\columnwidth]{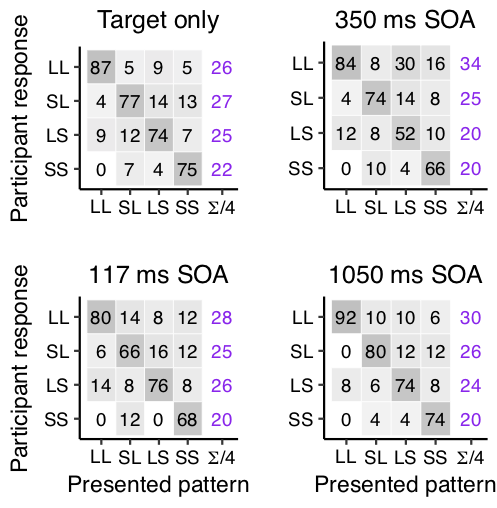}
  \caption{%
    Confusion matrices for Experiment 2B. Stimuli are columns, responses rows. Accuracy is in \%.
		Finger 2 is index finger, 3 is middle finger, 4 is ring, 5 is pinky.
    \label{fig:2b-confusion}%
  }
\end{figure}

\paragraph{Pattern.}
There was a significant main effect of Pattern ($\chi^{2}(3)=23.8, p<.001$). Performance with the LL pattern was better than the other three ($M_{LL}=86.7\%$ vs. $M_{SL}=76.1\%$, $M_{LS}=72.7\%$ \& $M_{SS}=74.2\%$, all $p<.001$); no other pairwise comparisons were significant (all $p>.32$).
Individual pattern performance is reported in Figure~\ref{fig:2b-confusion}.

\paragraph{Interactions.}
Neither the Trial Type $\times$ Session Time interaction ($\chi^{2}(3)=3.16, p=.44$) nor the Trial Type $\times$ Pattern interaction ($\chi^{2}(9)=16.6, p=.48$) reached significance.

\subsection{Discussion}
Auditory surprise distractors impaired performance rather modestly, and this effect was only apparent for the 350 ms SOA. As we observed in Experiment 2A, participants became better at the task with practice and showed a bias towards reporting patterns that involved a long second stimulus. Also like Experiment 2A, fewer participants were excluded for low baseline peformance (12\% in Experiments 2A and 2B, as compared to 25\% in Experiments 1A and 1B). This better performance may reflect more reliable discrimination in this task across users, which is important for HCI applications. Despite such better group performance, however, the surprise distraction effects were still significant and clear. Additional comparisons and a general discussion of all four experiments follows in the next section.

\section{General Discussion: Considerations and Recommendations for HCI Design}
A relatively rare, unexpected vibration or sound can disrupt subsequent haptic processing. Such distraction can occur even when the two sensory inputs are separated by space, time, or modality. Furthermore, the second input can be disrupted even when the first could have served as a helpful alert or notification. Although the degree of potential distraction will vary depending on the specifics of the interface in question, we argue that testing for such effects is critical for the optimal design of HCI experiments and applications.

In the present study, which included four different experiments using simulated smartrings, we showed that a relatively rare and unexpected stimulus can disrupt vibrotactile target detection and pattern discrimination tasks. This deleterious effect was present for both auditory and vibrotactile surprise distrators, though the latter caused more pronounced deficits. When each surprise distractor preceded the target by 117 or 350 ms, the deficits persisted throughout the entire session (>20 surprise distractors). By contrast, vibrotactile distractors that preceded targets by 1050 ms led to a deficit early in the session and a beneficial alert later. Finally, target detection was best on the index and middle fingers~\cite{VegaBermudez2001,King2010,Sathian1996}, whereas the long-long vibration pattern was selected most often, perhaps representing both stimulus discriminability and response bias~\cite{Kaaresoja2005}.

We believe that the results of our experiments have implications for multiple HCI contexts. Below we discuss their relevance to haptic information delivery, including via smartrings and other devices. The discussion also includes design guidelines, concrete suggestions, and additional ideas for future developments.

\subsection{Use Distractors, not Treadmills, for Ecological Validity!}
HCI research often requires ecologically valid experiments, reflecting device performance during use in real-world scenarios. For example, treadmills simulate the walking or moderate exercise in which users may be engaged when they receive a vibrotactile alert~\cite{Roumen2015,Pokering}. However, the effect of moderate physical activity on vibrotactile recognition is modest at best~\cite{CindyAVI2018}. In contrast, the present experiments show that distractors, which are certainly encountered during real-world device use, can have a potentially large effect on vibrotactile pattern recognition.
\paragraph{Recommendation.} Researchers looking to improve the ecological validity of their experiments should consider using sounds and vibrotactile distractors--and then evaluating their effects.

\subsection{Optimal Fingers for Devices}
Fingers have different sensitivity to haptic stimuli~\cite{VegaBermudez2001,King2010,Sathian1996}. Consistent with previous work, we found the best fingers to use were the index and middle fingers. Spatial confusion was highest between adjacent fingers, however, and the degree of such confusion may reduce the benefits of receiving vibrotactile feedback on multiple fingers.
In Experiment 1, the index and middle fingers had the highest performance during target-only trials and appeared to suffer less from surprise distractors, at least vibrotactile ones (Experiment 1A). These fingers' reduced susceptibility to distraction may simply reflect their higher performance baseline. Alternatively, intrinsic differences across the fingers or even the placement of the wrist motor may affect surprise distractor effects~\cite{Laksh2015}.

Regardless of the receiving finger, accurate transmission is unsurprisingly helped by using higher-intensity vibrations. Indeed, in Experiment 2, we found that our Long-Long pattern was more easily recognized that the other patterns.
Since all four patterns have the same overall duration, we hypothesize that the key factor here is the intensity of the vibration, as our long pulses had a higher intensity than the short ones (see Figure~\ref{fig:2a-apparatus}).

\paragraph{Recommendation.} For applications with a single finger-based device, use the index or middle finger (as opposed to the standard ring finger).
\paragraph{Recommendation.} For applications with multiple devices, consider further spatial separation to reduce confusion between adjacent locations.
\paragraph{Recommendation.} Use higher-intensity vibrotactile stimulation to increase perceptual certainty, as that may provide some buffer against distraction.

\subsection{Active Compensation for Unexpected Environmental Events}
In real-world scenarios, surprise distractors will often be stimuli in the environment.
For example, a brief loud sound, a buzzing phone, a chirping messaging notification, or an intermittent vibration on the train could disrupt an ongoing task in any sensory modality.
To maximize user performance, a system conveying information to the user could detect distracting environmental stimuli with sensors and then delay presenting crucial information.
Our results show that even a delay of a few seconds (appropriate for many types of notifications) could drastically improve information transfer.
Note that although such a delay is relatively short, it is still considerably longer than the approximately 100 ms delay that would be recommended based on perceptual (non-attentional) considerations alone.
As we discuss further below, the user’s attentional focus appears to be the limiting factor~\cite{Craig1995}.

Instead of detecting specific distracting events, the system could instead detect when the user is in a taxing environment or a taxing attentional state~\cite{Gevins2003,Loft2007,Murphy2016,Murphy2018,Baldwin2012,Marsalia2016,Qian2014}. In such cases, information that is important but does not require immediate action (e.g. an e-mail notification) could be presented at a slightly later time. Such strategic timing would ensure that the target vibration was not missed, while simultaneously preventing that stimulus from becoming a task-disrupting surprise distractor itself.

\paragraph{Recommendation.} Design and deploy systems that detect environmental events and delay presenting critical information accordingly.
\paragraph{Recommendation.} Design and deploy systems that detect user stress (potential or actual) and delay presenting critical information accordingly.

\subsection{Intelligent Dynamic Scheduling}
Some systems (e.g. computers or mobile phones, including their peripheral devices) control multiple streams of information across multiple modalities. For such systems, dynamic scheduling can be used so that the user is never presented with two inputs in close succession. Such dynamic scheduling could be achieved within specific software applications or, better still, at the level of the operating system.

Dynamic scheduling would allow alerting stimuli, which are intended to be helpful, to avoid become surprise distractors for other channels of information. Another consideration with such alerts is that training may be required to avoid negative effects. For example, we found that vibrotactile surprise distractors still impaired pattern discrimination after 1050 ms (Experiment 2A), but this initial deficit turned into a benefit after more exposure with the vibrotactile "alert". Indeed, it signalled that additional important information would be presented soon, thereby allowing participants to prepare for immiment pattern recognition~\cite{Marsalia2016}. In contrast, users did not, and perhaps could not, use the surprise distractors to improve performance at shorter SOA values (e.g. 117 or 350 ms). As such, an optimal SOA value for aiding performance is likely closer to 1050 ms, though an optimal value for aiding performance remains to be determined.

The effects of auditory surprise distractors on our vibrotactile tasks have three features that are different from their vibrotactile surprise distractor counterparts. These differences have important implications for dynamic scheduling across modalities. First, the auditory surprise distractor effects were consistently modest, when present at all. Such smaller effects are consistent with other studies of cross-modal attention distraction~\cite{Murphy2018,Murphy2016,Hillstrom2002,Dellacqua2006,Soto2002}. Second, there was little evidence of effects in the 1050 ms SOA, either positive or negative. Third, the effects in the 117 ms SOA condition--when the vibrotactile surprise distractor effects were largest--were not significant for the auditory surprise distractors. It is notable that the timecourse of auditory distraction effects appeared to be different from the vibrotactile surprise distractors' effects, as these distractor types have similar cross-modal effects on visual tasks~\cite{Ljungberg2012}. Regardless, our results suggest that auditory notifications may be used with fewer negative effects during haptic tasks, with only a time around 350 ms SOA being critical for performance. As with the vibrotactile surprise distractors, however, the best and worst SOA values for presenting auditory information remain unknown. Characterizing these times is one aim of future research.

\paragraph{Recommendation.} Expect that events should be scheduled so that they are separated by at least one second.
\paragraph{Recommendation.} Use rare, unexpected alerts with care. Test whether using a different alert modality can help, or if repeated exposure to the alert renders it more consistently beneficial.
\paragraph{Recommendation.} Avoid spliting feedback between multiple devices, even if across modalities, unless their interactions have been characterized and their relative timings can be centrally controlled.

\subsection{Case Study: Multimodal Feedback for Digital Maps}
Digital Maps allow visually impaired users to browse through maps and retrieve spatial information by moving their hands and fingers on a flat surface. Digital Maps use both audio and vibrotactile feedback to convey information to the user. Audio feedback is used to indicate the name of places, while vibrotactile feedback represents borders between spatial divisions. Recent investigation on combining both types of feedback showed that users may potentially be able to navigate digital maps with two hands and that bilateral audio feedback (i.e. coming from two distinct locations) may make navigation faster~\cite{SandraInteract2019}.
However, our results from this work have two implications for such scenarios:
\begin{enumerate}
  \item Providing vibrotactile feedback on the fingers of the same hand may be rather confusing in some cases, as participants may be unable to tell on which finger the feedback is sent. This potential confusion should be tested and then mitigated.
  \item Intermittent audio feedback may also have a negative effect on vibrotactile feedback. A suggestion would be to delay sending vibrotactile feedback for at least 350 milliseconds after sending audio feedback. Again, specific testing and mitigation for this application is appropriate.
\end{enumerate}

\section{General Discussion: To Psychological Theory and Back Again}
Our work can be evaluated within the context of psychological theory, and it also contributes to that literature. We discuss the contex and contributions below, as we believe that many related ideas from psychological experimentation could be usefully applied to HCI matters in the future.

\subsection{Attentional Capture}
The present results are relevant to fundamental questions in the cognitive psychology of attention and its control. Although psychological research and theory focus primarily on attention's beneficial effects, it has a "dark side": Unattended items are frequently missed~\cite{Chun2002}. In our experiments, detection failures were found for both target detection and pattern discrimination, and both within and across modalities. One possibility is that the observed deficits were due to surprise capture, wherein relatively rare and unexpected distractors grab attention in a stimulus-driven manner~\cite{Egeth1997,Asplund2010,Horstmann2015}. Another possibility is that the deficits were due to contingent capture, wherein the distracting item captures attention because it contains a target-defining feature~\cite{Folk1992,Folk2002}. For example, the vibrotactile surprise distractors in Experiment 1 had target-defining features (strong vibrations). Conversely, the auditory surprise distractors had no such shared features and so more clearly represent surprise capture. Contingent and surprise capture are also not mutually exclusive, and both effects could occur within a single paradigm. Future research will be required to distinguish their contributions, though we highlight relevant evidence in our results below.

Importantly, the deficits are likely due to attentional capture, not perceptual effects such as masking~\cite{Craig1995,Hillstrom2002,Craig1982}. Foremost, tactile masking effects are typically found for stimulus separations below 100 ms, not our 150 ms (Experiment 1A) or 233 ms (350 ms SOA condition in Experiment 2A) intervals between surprise distractor offset and target onset~\cite{Craig1982,Enriquez2008,Weisenberger1982,Gallace2006}. In addition, masking effects for stimulus pairs that are presented asynchronously are typically found for backwards masking, not the foreward masking reported here~\cite{Craig1982,Craig1995,Enriquez2008}. Although perceptual masking has been demonstrated across skin sites~\cite{Craig1995,Enriquez2008,Weisenberger1982}, including wrist-to-finger~\cite{Laksh2015} and even across large distances such as one forearm to another~\cite{Damour2014}, such effects involve simultaneous presentation of the stimuli in question. We therefore argue that our interference effects involve late processes involved in attentional orienting~\cite{Craig1995}. Our cross-modal effects also implicate late processes and are inconsistent with a masking account.

\subsection{Effect Timecourse Within Trials}
For both auditory and visual attention, attentional capture effects tend to be the greatest 200-400 ms after the inducing stimulus' onset~\cite{Asplund2010,Horstmann2015,Folk1992,Folk2002,Vachon2017}. Similar effect timing appears to apply for crossmodal surprise distractor deficits, including from vibrotactile stimuli~\cite{Parmentier2011,Ljungberg2012}, as well as in similar temporal attention paradigms such as the visual attentional blink~\cite{Dux2009}. Our auditory-vibrotactile cross-modal effects (Experiment 2B) appears to develop during each trial with a broadly comparable timecourse to these other studies as well. In contrast, our vibrotactile effects were strong even at 117 ms SOA. Vibrotactile attentional effects may differ in their timecourse; indeed, haptic attentional blinks appear to involve more pronounced short-SOA (100-200 ms) effects~\cite{Hillstrom2002,Dellacqua2006}.

The modalities also differ in their spatial aspects, which may contribute to differences in the the magnitude of the effect and its timecourse. Tactile attentional limitations, however, appear to be similar for successive stimuli at different locations or successive stimuli at the same location~\cite{Hillstrom2002,Dellacqua2006,Murphy2018}. Surprise and contingent capture effects are also evident with or without such spatial components~\cite{Asplund2010b,Horstmann2015,Folk1992,Folk2002}. Further investigation will be required to understand the differences across surprise distractor modality, particularly with regards to the effect's timecourse within each trial.

\subsection{Effect Timecourse Across the Session}
In the visual domain, surprise effects are evidenced for only a few trials~\cite{Asplund2010,Horstmann2015}. Auditory attentional capture appears to last longer, yet the effect still eventually habituates~\cite{Vachon2017}. Vibrotactile surprise distractors might be expected to have a timecourse similar to their auditory counterparts, as has been empirically demonstrated~\cite{Parmentier2011,Ljungberg2012}. Since we observed little change in our surprise distractor effects across each session, capture effects remained robust and/or participants did not learn to use the surprise distractor as a beneficial cue. One possibility is that our surprise effects habituated slowly, perhaps outside the time window of the experimental session. Another possibility is that the attentional capture deficits were due, at least in part, to contingent capture effects~\cite{Folk1992,Folk2002}. These effects have not been reported to habituate. An important exception is found in Experiment 2A, where there was significant habituation for the longest SOA (1050 ms). We surmise, however, that this habituation was due to the voluntary redirection of attention to facilitate target detection, a use of the surprise distractor that emerged only after experience. Note that such experience was identical for shorter SOAs as well, though there was no habituation of the distracting effects in those conditions.
\section{Limitations}
Our experiments provided informative results and clear effects, but they also contain some limitations.
First, surprise trials have rare and unexpected events by definition, so their numbers had to be relatively low to keep the experimental session at a reasonable length. We relied on previous psychological work to find a high surprise trial frequency (around 15\%) that still produced reliable effects~\cite{Asplund2010,Horstmann2015,Yamaguchi1991}. In addition, we recruited a large sample (>100 across 4 experiments) to increase our statistical power. Nevertheless, some effects--notably the expected habituation of the deficits--were hinted at in many experiments, suggesting that additional power is necessary to determine whether they are reliably present.

Second, we used off-the-shelf vibration motors, similar to motors in commercial products such as mobile phones. Such inexpensive motors have some control difficulties; indeed, our short pulse intensity was considerably lower than our long pulse intensity in Experiment 2. This intensity difference may explain the superior performance with the long-long pattern. Nevertheless, participant performance was in line with related work in HCI~\cite{Alvina2015,Lee2010,Yatani2009,Yatani2012}.

Third, participants in our study wore the wristband-embedded motor (mimicing a smartwatch) and finger devices (mimicing smart rings) on the same hand, which may not reflect the typical real-world configuration. As such, our setup represented a worst-case scenario in this regard.

Finally, we used a narrow range of vibrotactile targets and surprise distractors, but the characteristics of such stimuli--as well as their perceived strength--may affect the degree of surprise distraction observed. Future experiments may investigate such effects by presenting vibrations to different sites that have been tuned so that they have similar perceived magnitudes. For our experiements, however, we elected to use similar target and surprise distractor intensities becasue off-the-shelf devices would presumably not have finely tuned vibration amplitudes. In addition, any such tuning with our motors would have yielded differences in amplitude, duration, and/or vibration frequency.

Nevertheless, our experiments and their results do have some relevance to questions of perceived stimulus intensity. Foremost, it is reasonable to assume that wrist stimulation is generally perceived as weaker than finger stimulation; the wrist-based vibration distractors were still effective at disrupting performance. Furthermore, the intensiies of the target patterns in Experiment 2 varied widely in relation to each other and to the surprise distractor. The surprise distractor was still effective as disrupting recognition of each pattern to approximately the same degree (see Figure~\ref{fig:2b-confusion}).

Relatedly, differences in perceived stimulus intensity complicate the comparison of surprise distraction magnitudes across vibrotactile and auditory distractors. Indeed, it was not our primary aim here to compare the magnitudes, though we note that the auditory stimuli used in Experiment 1B can induce much larger unimodal deficits than those observed in the present study~\cite{Obana2019}. Although differences in magnitude across the modalities should be interpreted with care, that each stimulus modality could induce a deficit is important, as are the different timecourses with which the observed deficits developed following each surprise distractor (Experiment 2).

Further exploration of the relationship between distraction effects and perceived stimulus properties (especially intensity) is warranted. Indeed, we are investigating such matters in ongoing experimentation in our laboratories.

\section{Conclusion}
A relatively rare, unexpected vibration or sound can disrupt subsequent haptic processing. In four experiments, we showed that our effects hold across spatial detection and pattern discrimination tasks, as well as across modalities (albeit with stronger intramodal vibrotactile effects). As such, our results show high internal validity, replicability, and generalizability. They are thus broadly applicable to vibrotactile tasks completed in the face of potential distraction, and they should be considered for the optimal design of HCI experiments and applications.

While our study primarily focuses on such negative effects, we also highlighted some potential benefits that deserve further study. In the future, we plan to investigate additional SOAs to find the optimal value for cueing benefits, thereby enabling alerts to be most effective at directing attention to forthcoming information delivery. We will also explore surprises that occur on the ring device itself, as both the positive (cueing) and negative (attentional capture) effects could be greater there. Finally, we hope to conduct field studies to understand the influence of environmental surprise distractors and their mitigation.

We conclude that distracting effects can be mitigated through active intervention, deliberate spacing of signals, and/or practice with potentially distracting stimuli so that they can become beneficial notifications over time. We therefore recommend that designers include such features in devices and user interfaces. Otherwise, even stimuli intended to be helpful can impair performance. Regardless of their intent, relatively rare and unexpected events can be disruptive to haptic tasks; it's all (mostly) in the timing.

\bibliographystyle{ACM-Reference-Format}
\balance
\bibliography{nokia}


\begin{thebibliography}{70}


\ifx \showCODEN    \undefined \def \showCODEN     #1{\unskip}     \fi
\ifx \showDOI      \undefined \def \showDOI       #1{#1}\fi
\ifx \showISBNx    \undefined \def \showISBNx     #1{\unskip}     \fi
\ifx \showISBNxiii \undefined \def \showISBNxiii  #1{\unskip}     \fi
\ifx \showISSN     \undefined \def \showISSN      #1{\unskip}     \fi
\ifx \showLCCN     \undefined \def \showLCCN      #1{\unskip}     \fi
\ifx \shownote     \undefined \def \shownote      #1{#1}          \fi
\ifx \showarticletitle \undefined \def \showarticletitle #1{#1}   \fi
\ifx \showURL      \undefined \def \showURL       {\relax}        \fi
\providecommand\bibfield[2]{#2}
\providecommand\bibinfo[2]{#2}
\providecommand\natexlab[1]{#1}
\providecommand\showeprint[2][]{arXiv:#2}

\bibitem[\protect\citeauthoryear{Alvina, Zhao, Perrault, Azh, Roumen, and
  Fjeld}{Alvina et~al\mbox{.}}{2015}]%
        {Alvina2015}
\bibfield{author}{\bibinfo{person}{Jessalyn Alvina}, \bibinfo{person}{Shengdong
  Zhao}, \bibinfo{person}{Simon~T. Perrault}, \bibinfo{person}{Maryam Azh},
  \bibinfo{person}{Thijs Roumen}, {and} \bibinfo{person}{Morten Fjeld}.}
  \bibinfo{year}{2015}\natexlab{}.
\newblock \showarticletitle{OmniVib: Towards Cross-body Spatiotemporal
  Vibrotactile Notifications for Mobile Phones}. In
  \bibinfo{booktitle}{\emph{Proceedings of the 33rd Annual ACM Conference on
  Human Factors in Computing Systems}} \emph{(\bibinfo{series}{CHI '15})}.
  \bibinfo{publisher}{ACM}, \bibinfo{address}{New York, NY, USA},
  \bibinfo{pages}{2487--2496}.
\newblock
\showISBNx{978-1-4503-3145-6}
\urldef\tempurl%
\url{https://doi.org/10.1145/2702123.2702341}
\showDOI{\tempurl}


\bibitem[\protect\citeauthoryear{Asplund, Fougnie, Zughni, Martin, and
  Marois}{Asplund et~al\mbox{.}}{2014}]%
        {Asplund2014}
\bibfield{author}{\bibinfo{person}{C.~L. Asplund}, \bibinfo{person}{D.
  Fougnie}, \bibinfo{person}{S. Zughni}, \bibinfo{person}{J.~W. Martin}, {and}
  \bibinfo{person}{R. Marois}.} \bibinfo{year}{2014}\natexlab{}.
\newblock \showarticletitle{The {Attentional} {Blink} {Reveals} the
  {Probabilistic} {Nature} of {Discrete} {Conscious} {Perception}}.
\newblock \bibinfo{journal}{\emph{Psychological Science}} \bibinfo{volume}{25},
  \bibinfo{number}{3} (\bibinfo{date}{March} \bibinfo{year}{2014}),
  \bibinfo{pages}{824--831}.
\newblock
\showISSN{0956-7976, 1467-9280}
\urldef\tempurl%
\url{https://doi.org/10.1177/0956797613513810}
\showDOI{\tempurl}


\bibitem[\protect\citeauthoryear{Asplund, Todd, Snyder, Gilbert, and
  Marois}{Asplund et~al\mbox{.}}{2010b}]%
        {Asplund2010}
\bibfield{author}{\bibinfo{person}{Christopher~L. Asplund},
  \bibinfo{person}{J.~Jay Todd}, \bibinfo{person}{A.~P. Snyder},
  \bibinfo{person}{Christopher~M. Gilbert}, {and} \bibinfo{person}{Ren{\'e}
  Marois}.} \bibinfo{year}{2010}\natexlab{b}.
\newblock \showarticletitle{Surprise-induced blindness: {A} stimulus-driven
  attentional limit to conscious perception.}
\newblock \bibinfo{journal}{\emph{Journal of Experimental Psychology: Human
  Perception and Performance}} \bibinfo{volume}{36}, \bibinfo{number}{6}
  (\bibinfo{year}{2010}), \bibinfo{pages}{1372--1381}.
\newblock
\showISSN{1939-1277, 0096-1523}
\urldef\tempurl%
\url{https://doi.org/10.1037/a0020551}
\showDOI{\tempurl}


\bibitem[\protect\citeauthoryear{Asplund, Todd, Snyder, and Marois}{Asplund
  et~al\mbox{.}}{2010a}]%
        {Asplund2010b}
\bibfield{author}{\bibinfo{person}{Christopher~L Asplund},
  \bibinfo{person}{J~Jay Todd}, \bibinfo{person}{Andy~P Snyder}, {and}
  \bibinfo{person}{Ren{\'e} Marois}.} \bibinfo{year}{2010}\natexlab{a}.
\newblock \showarticletitle{A central role for the lateral prefrontal cortex in
  goal-directed and stimulus-driven attention}.
\newblock \bibinfo{journal}{\emph{Nature Neuroscience}} \bibinfo{volume}{13},
  \bibinfo{number}{4} (\bibinfo{date}{April} \bibinfo{year}{2010}),
  \bibinfo{pages}{507--512}.
\newblock
\showISSN{1097-6256, 1546-1726}
\urldef\tempurl%
\url{https://doi.org/10.1038/nn.2509}
\showDOI{\tempurl}


\bibitem[\protect\citeauthoryear{Baldwin, Spence, Bliss, Brill, Wogalter,
  Mayhorn, and Ferris}{Baldwin et~al\mbox{.}}{2012}]%
        {Baldwin2012}
\bibfield{author}{\bibinfo{person}{Carryl~L. Baldwin}, \bibinfo{person}{Charles
  Spence}, \bibinfo{person}{James~P. Bliss}, \bibinfo{person}{J.~Christopher
  Brill}, \bibinfo{person}{Michael~S. Wogalter},
  \bibinfo{person}{Christopher~B. Mayhorn}, {and} \bibinfo{person}{Thomas~K.
  Ferris}.} \bibinfo{year}{2012}\natexlab{}.
\newblock \showarticletitle{Multimodal Cueing: The Relative Benefits of the
  Auditory, Visual, and Tactile Channels in Complex Environments}.
\newblock \bibinfo{journal}{\emph{Proceedings of the Human Factors and
  Ergonomics Society Annual Meeting}} \bibinfo{volume}{56}, \bibinfo{number}{1}
  (\bibinfo{year}{2012}), \bibinfo{pages}{1431--1435}.
\newblock
\urldef\tempurl%
\url{https://doi.org/10.1177/1071181312561404}
\showDOI{\tempurl}
\showeprint{https://doi.org/10.1177/1071181312561404}


\bibitem[\protect\citeauthoryear{Bardot, Serrano, Perrault, Zhao, and
  Jouffrais}{Bardot et~al\mbox{.}}{2019}]%
        {SandraInteract2019}
\bibfield{author}{\bibinfo{person}{Sandra Bardot}, \bibinfo{person}{Marcos
  Serrano}, \bibinfo{person}{Simon Perrault}, \bibinfo{person}{Shengdong Zhao},
  {and} \bibinfo{person}{Christophe Jouffrais}.}
  \bibinfo{year}{2019}\natexlab{}.
\newblock \showarticletitle{Investigating Feedback for Two-Handed Exploration
  of Digital Maps Without Vision}. In \bibinfo{booktitle}{\emph{Human-Computer
  Interaction -- INTERACT 2019}}, \bibfield{editor}{\bibinfo{person}{David
  Lamas}, \bibinfo{person}{Fernando Loizides}, \bibinfo{person}{Lennart Nacke},
  \bibinfo{person}{Helen Petrie}, \bibinfo{person}{Marco Winckler}, {and}
  \bibinfo{person}{Panayiotis Zaphiris}} (Eds.). \bibinfo{publisher}{Springer
  International Publishing}, \bibinfo{address}{Cham},
  \bibinfo{pages}{305--324}.
\newblock
\showISBNx{978-3-030-29381-9}


\bibitem[\protect\citeauthoryear{Barr, Levy, Scheepers, and Tily}{Barr
  et~al\mbox{.}}{2013}]%
        {Barr2013}
\bibfield{author}{\bibinfo{person}{Dale~J. Barr}, \bibinfo{person}{Roger Levy},
  \bibinfo{person}{Christoph Scheepers}, {and} \bibinfo{person}{Harry~J.
  Tily}.} \bibinfo{year}{2013}\natexlab{}.
\newblock \showarticletitle{Random effects structure for confirmatory
  hypothesis testing: {Keep} it maximal}.
\newblock \bibinfo{journal}{\emph{Journal of Memory and Language}}
  \bibinfo{volume}{68}, \bibinfo{number}{3} (\bibinfo{date}{April}
  \bibinfo{year}{2013}), \bibinfo{pages}{255--278}.
\newblock
\showISSN{0749596X}
\urldef\tempurl%
\url{https://doi.org/10.1016/j.jml.2012.11.001}
\showDOI{\tempurl}


\bibitem[\protect\citeauthoryear{Bates, M{\"a}chler, Bolker, and Walker}{Bates
  et~al\mbox{.}}{2015}]%
        {Bates2015}
\bibfield{author}{\bibinfo{person}{Douglas Bates}, \bibinfo{person}{Martin
  M{\"a}chler}, \bibinfo{person}{Ben Bolker}, {and} \bibinfo{person}{Steve
  Walker}.} \bibinfo{year}{2015}\natexlab{}.
\newblock \showarticletitle{Fitting {Linear} {Mixed}-{Effects} {Models} {Using}
  lme4}.
\newblock \bibinfo{journal}{\emph{Journal of Statistical Software}}
  \bibinfo{volume}{67}, \bibinfo{number}{1} (\bibinfo{year}{2015}),
  \bibinfo{pages}{1--48}.
\newblock
\urldef\tempurl%
\url{https://doi.org/10.18637/jss.v067.i01}
\showDOI{\tempurl}


\bibitem[\protect\citeauthoryear{Benjamini and Hochberg}{Benjamini and
  Hochberg}{1995}]%
        {Benjamini1995}
\bibfield{author}{\bibinfo{person}{Yoav Benjamini} {and} \bibinfo{person}{Yosef
  Hochberg}.} \bibinfo{year}{1995}\natexlab{}.
\newblock \showarticletitle{Controlling the {False} {Discovery} {Rate}: {A}
  {Practical} and {Powerful} {Approach} to {Multiple} {Testing}}.
\newblock \bibinfo{journal}{\emph{Journal of the Royal Statistical Society.
  Series B (Methodological)}} \bibinfo{volume}{57}, \bibinfo{number}{1}
  (\bibinfo{year}{1995}), \bibinfo{pages}{289--300}.
\newblock


\bibitem[\protect\citeauthoryear{Brouwer and Van~Erp}{Brouwer and
  Van~Erp}{2010}]%
        {Brouwer2010}
\bibfield{author}{\bibinfo{person}{Anne-Marie Brouwer} {and}
  \bibinfo{person}{Jan Van~Erp}.} \bibinfo{year}{2010}\natexlab{}.
\newblock \showarticletitle{A tactile P300 brain-computer interface}.
\newblock \bibinfo{journal}{\emph{Frontiers in Neuroscience}}
  \bibinfo{volume}{4} (\bibinfo{year}{2010}), \bibinfo{pages}{19}.
\newblock
\showISSN{1662-453X}
\urldef\tempurl%
\url{https://doi.org/10.3389/fnins.2010.00019}
\showDOI{\tempurl}


\bibitem[\protect\citeauthoryear{Carrasco, Ling, and Read}{Carrasco
  et~al\mbox{.}}{2004}]%
        {Carrasco2004}
\bibfield{author}{\bibinfo{person}{Marisa Carrasco}, \bibinfo{person}{Sam
  Ling}, {and} \bibinfo{person}{Sarah Read}.} \bibinfo{year}{2004}\natexlab{}.
\newblock \showarticletitle{Attention alters appearance}.
\newblock \bibinfo{journal}{\emph{Nature Neuroscience}} \bibinfo{volume}{7},
  \bibinfo{number}{3} (\bibinfo{date}{March} \bibinfo{year}{2004}),
  \bibinfo{pages}{308--313}.
\newblock
\showISSN{1097-6256}
\urldef\tempurl%
\url{https://doi.org/10.1038/nn1194}
\showDOI{\tempurl}


\bibitem[\protect\citeauthoryear{Chen and Terrence}{Chen and Terrence}{2008}]%
        {Chen2008}
\bibfield{author}{\bibinfo{person}{J.~Y.C. Chen} {and} \bibinfo{person}{P.~I.
  Terrence}.} \bibinfo{year}{2008}\natexlab{}.
\newblock \showarticletitle{Effects of tactile cueing on concurrent performance
  of military and robotics tasks in a simulated multitasking environment}.
\newblock \bibinfo{journal}{\emph{Ergonomics}} \bibinfo{volume}{51},
  \bibinfo{number}{8} (\bibinfo{year}{2008}), \bibinfo{pages}{1137--1152}.
\newblock
\urldef\tempurl%
\url{https://doi.org/10.1080/00140130802030706}
\showDOI{\tempurl}
\showeprint{https://doi.org/10.1080/00140130802030706}
\newblock
\shownote{PMID: 18608472.}


\bibitem[\protect\citeauthoryear{Chen, Perrault, Roy, and Wyse}{Chen
  et~al\mbox{.}}{2018}]%
        {CindyAVI2018}
\bibfield{author}{\bibinfo{person}{Qin Chen}, \bibinfo{person}{Simon~T.
  Perrault}, \bibinfo{person}{Quentin Roy}, {and} \bibinfo{person}{Lonce
  Wyse}.} \bibinfo{year}{2018}\natexlab{}.
\newblock \showarticletitle{Effect of Temporality, Physical Activity and
  Cognitive Load on Spatiotemporal Vibrotactile Pattern Recognition}. In
  \bibinfo{booktitle}{\emph{Proceedings of the 2018 International Conference on
  Advanced Visual Interfaces}} \emph{(\bibinfo{series}{AVI '18})}.
  \bibinfo{publisher}{Association for Computing Machinery},
  \bibinfo{address}{New York, NY, USA}, Article \bibinfo{articleno}{Article
  25}, \bibinfo{numpages}{9}~pages.
\newblock
\showISBNx{9781450356169}
\urldef\tempurl%
\url{https://doi.org/10.1145/3206505.3206511}
\showDOI{\tempurl}


\bibitem[\protect\citeauthoryear{Cholewiak and Craig}{Cholewiak and
  Craig}{1984}]%
        {Cholewiak1984}
\bibfield{author}{\bibinfo{person}{Roger~W Cholewiak} {and}
  \bibinfo{person}{James~C Craig}.} \bibinfo{year}{1984}\natexlab{}.
\newblock \showarticletitle{Vibrotactile pattern recognition and discrimination
  at several body sites}.
\newblock \bibinfo{journal}{\emph{Perception \& psychophysics}}
  \bibinfo{volume}{35}, \bibinfo{number}{6} (\bibinfo{year}{1984}),
  \bibinfo{pages}{503--514}.
\newblock


\bibitem[\protect\citeauthoryear{Chun and Marois}{Chun and Marois}{2002}]%
        {Chun2002}
\bibfield{author}{\bibinfo{person}{Marvin~M Chun} {and}
  \bibinfo{person}{Ren{\'e} Marois}.} \bibinfo{year}{2002}\natexlab{}.
\newblock \showarticletitle{The dark side of visual attention}.
\newblock \bibinfo{journal}{\emph{Current Opinion in Neurobiology}}
  \bibinfo{volume}{12}, \bibinfo{number}{2} (\bibinfo{date}{April}
  \bibinfo{year}{2002}), \bibinfo{pages}{184--189}.
\newblock
\showISSN{09594388}
\urldef\tempurl%
\url{https://doi.org/10.1016/S0959-4388(02)00309-4}
\showDOI{\tempurl}
\newblock
\shownote{00153.}


\bibitem[\protect\citeauthoryear{Craig}{Craig}{1982}]%
        {Craig1982}
\bibfield{author}{\bibinfo{person}{James~C. Craig}.}
  \bibinfo{year}{1982}\natexlab{}.
\newblock \showarticletitle{Vibrotactile masking: {A} comparison of energy and
  pattern maskers}.
\newblock \bibinfo{journal}{\emph{Perception \& Psychophysics}}
  \bibinfo{volume}{31}, \bibinfo{number}{6} (\bibinfo{date}{Nov.}
  \bibinfo{year}{1982}), \bibinfo{pages}{523--529}.
\newblock
\showISSN{0031-5117, 1532-5962}
\urldef\tempurl%
\url{https://doi.org/10.3758/BF03204184}
\showDOI{\tempurl}


\bibitem[\protect\citeauthoryear{Craig and Evans}{Craig and Evans}{1995}]%
        {Craig1995}
\bibfield{author}{\bibinfo{person}{James~C. Craig} {and}
  \bibinfo{person}{Paul~M. Evans}.} \bibinfo{year}{1995}\natexlab{}.
\newblock \showarticletitle{Tactile selective attention and temporal masking}.
\newblock \bibinfo{journal}{\emph{Attention, Perception, \& Psychophysics}}
  \bibinfo{volume}{57}, \bibinfo{number}{4} (\bibinfo{year}{1995}),
  \bibinfo{pages}{511--518}.
\newblock
\urldef\tempurl%
\url{http://www.springerlink.com/index/272215101PX7Q387.pdf}
\showURL{%
\tempurl}


\bibitem[\protect\citeauthoryear{D'Amour and Harris}{D'Amour and
  Harris}{2014}]%
        {Damour2014}
\bibfield{author}{\bibinfo{person}{Sarah D'Amour} {and}
  \bibinfo{person}{Laurence~R. Harris}.} \bibinfo{year}{2014}\natexlab{}.
\newblock \showarticletitle{Contralateral tactile masking between forearms}.
\newblock \bibinfo{journal}{\emph{Experimental Brain Research}}
  \bibinfo{volume}{232}, \bibinfo{number}{3} (\bibinfo{date}{March}
  \bibinfo{year}{2014}), \bibinfo{pages}{821--826}.
\newblock
\showISSN{0014-4819, 1432-1106}
\urldef\tempurl%
\url{https://doi.org/10.1007/s00221-013-3791-y}
\showDOI{\tempurl}


\bibitem[\protect\citeauthoryear{Dell'Acqua, Jolic{\oe}ur, Sessa, and
  Turatto}{Dell'Acqua et~al\mbox{.}}{2006}]%
        {Dellacqua2006}
\bibfield{author}{\bibinfo{person}{R. Dell'Acqua}, \bibinfo{person}{P.
  Jolic{\oe}ur}, \bibinfo{person}{P. Sessa}, {and} \bibinfo{person}{M.
  Turatto}.} \bibinfo{year}{2006}\natexlab{}.
\newblock \showarticletitle{Attentional blink and selection in the tactile
  domain}.
\newblock \bibinfo{journal}{\emph{European Journal of Cognitive Psychology}}
  \bibinfo{volume}{18}, \bibinfo{number}{4} (\bibinfo{date}{July}
  \bibinfo{year}{2006}), \bibinfo{pages}{537--559}.
\newblock
\showISSN{0954-1446, 1464-0635}
\urldef\tempurl%
\url{https://doi.org/10.1080/09541440500423186}
\showDOI{\tempurl}


\bibitem[\protect\citeauthoryear{Donchin, Spencer, and Wijesinghe}{Donchin
  et~al\mbox{.}}{2000}]%
        {Donchin2000}
\bibfield{author}{\bibinfo{person}{E. Donchin}, \bibinfo{person}{K.M. Spencer},
  {and} \bibinfo{person}{R. Wijesinghe}.} \bibinfo{year}{2000}\natexlab{}.
\newblock \showarticletitle{The mental prosthesis: assessing the speed of a
  {P}300-based brain-computer interface}.
\newblock \bibinfo{journal}{\emph{IEEE Transactions on Rehabilitation
  Engineering}} \bibinfo{volume}{8}, \bibinfo{number}{2} (\bibinfo{date}{June}
  \bibinfo{year}{2000}), \bibinfo{pages}{174--179}.
\newblock
\showISSN{10636528}
\urldef\tempurl%
\url{https://doi.org/10.1109/86.847808}
\showDOI{\tempurl}


\bibitem[\protect\citeauthoryear{Dux and Marois}{Dux and Marois}{2009}]%
        {Dux2009}
\bibfield{author}{\bibinfo{person}{P.~E. Dux} {and} \bibinfo{person}{R.
  Marois}.} \bibinfo{year}{2009}\natexlab{}.
\newblock \showarticletitle{The attentional blink: {A} review of data and
  theory}.
\newblock \bibinfo{journal}{\emph{Attention, Perception \& Psychophysics}}
  \bibinfo{volume}{71}, \bibinfo{number}{8} (\bibinfo{date}{Nov.}
  \bibinfo{year}{2009}), \bibinfo{pages}{1683--1700}.
\newblock
\showISSN{1943-3921, 1943-393X}
\urldef\tempurl%
\url{https://doi.org/10.3758/APP.71.8.1683}
\showDOI{\tempurl}


\bibitem[\protect\citeauthoryear{Egeth and Yantis}{Egeth and Yantis}{1997}]%
        {Egeth1997}
\bibfield{author}{\bibinfo{person}{Howard~E. Egeth} {and}
  \bibinfo{person}{Steven Yantis}.} \bibinfo{year}{1997}\natexlab{}.
\newblock \showarticletitle{Visual attention: {Control}, representation, and
  time course}.
\newblock \bibinfo{journal}{\emph{Annual review of psychology}}
  \bibinfo{volume}{48}, \bibinfo{number}{1} (\bibinfo{year}{1997}),
  \bibinfo{pages}{269--297}.
\newblock
\urldef\tempurl%
\url{http://www.annualreviews.org/doi/abs/10.1146/annurev.psych.48.1.269}
\showURL{%
\tempurl}


\bibitem[\protect\citeauthoryear{Enriquez and MacLean}{Enriquez and
  MacLean}{2008}]%
        {Enriquez2008}
\bibfield{author}{\bibinfo{person}{Mario Enriquez} {and}
  \bibinfo{person}{Karon~E. MacLean}.} \bibinfo{year}{2008}\natexlab{}.
\newblock \showarticletitle{Backward and common-onset masking of vibrotactile
  stimuli}.
\newblock \bibinfo{journal}{\emph{Brain Research Bulletin}}
  \bibinfo{volume}{75}, \bibinfo{number}{6} (\bibinfo{date}{April}
  \bibinfo{year}{2008}), \bibinfo{pages}{761--769}.
\newblock
\showISSN{03619230}
\urldef\tempurl%
\url{https://doi.org/10.1016/j.brainresbull.2008.01.018}
\showDOI{\tempurl}


\bibitem[\protect\citeauthoryear{F.~Font}{F.~Font}{2013}]%
        {Font2013}
\bibfield{author}{\bibinfo{person}{X.~Serra F.~Font, G.~Roma}.}
  \bibinfo{year}{2013}\natexlab{}.
\newblock \showarticletitle{Freesound technical demo}.
\newblock \bibinfo{journal}{\emph{Proceedings of the 21st ACM International
  Conference on Multimedia}} (\bibinfo{year}{2013}), \bibinfo{pages}{411--412}.
\newblock


\bibitem[\protect\citeauthoryear{Folk, Leber, and Egeth}{Folk
  et~al\mbox{.}}{2002}]%
        {Folk2002}
\bibfield{author}{\bibinfo{person}{Charles~L. Folk}, \bibinfo{person}{Andrew~B.
  Leber}, {and} \bibinfo{person}{Howard~E. Egeth}.}
  \bibinfo{year}{2002}\natexlab{}.
\newblock \showarticletitle{Made you blink! {Contingent} attentional capture
  produces a spatial blink}.
\newblock \bibinfo{journal}{\emph{Perception \& psychophysics}}
  \bibinfo{volume}{64}, \bibinfo{number}{5} (\bibinfo{year}{2002}),
  \bibinfo{pages}{741--753}.
\newblock
\urldef\tempurl%
\url{http://link.springer.com/article/10.3758/BF03194741}
\showURL{%
\tempurl}


\bibitem[\protect\citeauthoryear{Folk, Remington, and Johnston}{Folk
  et~al\mbox{.}}{1992}]%
        {Folk1992}
\bibfield{author}{\bibinfo{person}{C.~L. Folk}, \bibinfo{person}{R.~W.
  Remington}, {and} \bibinfo{person}{J.~C. Johnston}.}
  \bibinfo{year}{1992}\natexlab{}.
\newblock \showarticletitle{Involuntary covert orienting is contingent on
  attentional control settings}.
\newblock \bibinfo{journal}{\emph{Journal of Experimental Psychology. Human
  Perception and Performance}} \bibinfo{volume}{18}, \bibinfo{number}{4}
  (\bibinfo{date}{Nov.} \bibinfo{year}{1992}), \bibinfo{pages}{1030--1044}.
\newblock
\showISSN{0096-1523}


\bibitem[\protect\citeauthoryear{Fox and Weisberg}{Fox and Weisberg}{2011}]%
        {Fox2011}
\bibfield{author}{\bibinfo{person}{J. Fox} {and} \bibinfo{person}{S.
  Weisberg}.} \bibinfo{year}{2011}\natexlab{}.
\newblock \bibinfo{booktitle}{\emph{An {R} {Companion} to {Applied}
  {Regression}}}.
\newblock \bibinfo{publisher}{SAGE}, \bibinfo{address}{Thousand Oaks, Calif.}
\newblock
\urldef\tempurl%
\url{http://socserv.socsci.mcmaster.ca/jfox/Books/Companion}
\showURL{%
\tempurl}


\bibitem[\protect\citeauthoryear{Gallace, Tan, and Spence}{Gallace
  et~al\mbox{.}}{2006}]%
        {Gallace2006}
\bibfield{author}{\bibinfo{person}{A. Gallace}, \bibinfo{person}{H.~Z. Tan},
  {and} \bibinfo{person}{C. Spence}.} \bibinfo{year}{2006}\natexlab{}.
\newblock \showarticletitle{The failure to detect tactile change: A tactile
  analogue of visual change blindness}.
\newblock \bibinfo{journal}{\emph{Psychonomic Bulletin \& Review}}
  \bibinfo{volume}{13}, \bibinfo{number}{2} (\bibinfo{year}{2006}),
  \bibinfo{pages}{300--303}.
\newblock


\bibitem[\protect\citeauthoryear{Gevins and Smith}{Gevins and Smith}{2003}]%
        {Gevins2003}
\bibfield{author}{\bibinfo{person}{Alan Gevins} {and}
  \bibinfo{person}{Michael~E. Smith}.} \bibinfo{year}{2003}\natexlab{}.
\newblock \showarticletitle{Neurophysiological measures of cognitive workload
  during human-computer interaction}.
\newblock \bibinfo{journal}{\emph{Theoretical Issues in Ergonomics Science}}
  \bibinfo{volume}{4}, \bibinfo{number}{1-2} (\bibinfo{year}{2003}),
  \bibinfo{pages}{113--131}.
\newblock
\urldef\tempurl%
\url{https://doi.org/10.1080/14639220210159717}
\showDOI{\tempurl}
\showeprint{https://doi.org/10.1080/14639220210159717}


\bibitem[\protect\citeauthoryear{Gibson and Craig}{Gibson and Craig}{2005}]%
        {Gibson2005}
\bibfield{author}{\bibinfo{person}{Gregory~O Gibson} {and}
  \bibinfo{person}{James~C Craig}.} \bibinfo{year}{2005}\natexlab{}.
\newblock \showarticletitle{Tactile spatial sensitivity and anisotropy}.
\newblock \bibinfo{journal}{\emph{Perception \& Psychophysics}}
  \bibinfo{volume}{67}, \bibinfo{number}{6} (\bibinfo{year}{2005}),
  \bibinfo{pages}{1061--1079}.
\newblock


\bibitem[\protect\citeauthoryear{Hillstrom, Shapiro, and Spence}{Hillstrom
  et~al\mbox{.}}{2002}]%
        {Hillstrom2002}
\bibfield{author}{\bibinfo{person}{Anne~P. Hillstrom},
  \bibinfo{person}{Kimron~L. Shapiro}, {and} \bibinfo{person}{Charles Spence}.}
  \bibinfo{year}{2002}\natexlab{}.
\newblock \showarticletitle{Attentional limitations in processing sequentially
  presented vibrotactile targets}.
\newblock \bibinfo{journal}{\emph{Attention, Perception, \& Psychophysics}}
  \bibinfo{volume}{64}, \bibinfo{number}{7} (\bibinfo{year}{2002}),
  \bibinfo{pages}{1068--1082}.
\newblock
\urldef\tempurl%
\url{http://www.springerlink.com/index/K1115663228P602Q.pdf}
\showURL{%
\tempurl}


\bibitem[\protect\citeauthoryear{Hoggan and Brewster}{Hoggan and
  Brewster}{2007}]%
        {Hoggan2007}
\bibfield{author}{\bibinfo{person}{Eve Hoggan} {and} \bibinfo{person}{Stephen
  Brewster}.} \bibinfo{year}{2007}\natexlab{}.
\newblock \showarticletitle{Designing audio and tactile crossmodal icons for
  mobile devices}. In \bibinfo{booktitle}{\emph{Proceedings of the 9th
  international conference on Multimodal interfaces}}. ACM,
  \bibinfo{pages}{162--169}.
\newblock


\bibitem[\protect\citeauthoryear{Hoggan, Kaaresoja, Laitinen, and
  Brewster}{Hoggan et~al\mbox{.}}{2008}]%
        {Hoggan2008}
\bibfield{author}{\bibinfo{person}{Eve Hoggan}, \bibinfo{person}{Topi
  Kaaresoja}, \bibinfo{person}{Pauli Laitinen}, {and} \bibinfo{person}{Stephen
  Brewster}.} \bibinfo{year}{2008}\natexlab{}.
\newblock \showarticletitle{Crossmodal congruence: the look, feel and sound of
  touchscreen widgets}. In \bibinfo{booktitle}{\emph{Proceedings of the 10th
  international conference on Multimodal interfaces}}. ACM,
  \bibinfo{pages}{157--164}.
\newblock


\bibitem[\protect\citeauthoryear{Horstmann}{Horstmann}{2015}]%
        {Horstmann2015}
\bibfield{author}{\bibinfo{person}{Gernot Horstmann}.}
  \bibinfo{year}{2015}\natexlab{}.
\newblock \showarticletitle{The surprise-attention link: a review: {The}
  surprise-attention link}.
\newblock \bibinfo{journal}{\emph{Annals of the New York Academy of Sciences}}
  \bibinfo{volume}{1339}, \bibinfo{number}{1} (\bibinfo{date}{March}
  \bibinfo{year}{2015}), \bibinfo{pages}{106--115}.
\newblock
\showISSN{00778923}
\urldef\tempurl%
\url{https://doi.org/10.1111/nyas.12679}
\showDOI{\tempurl}


\bibitem[\protect\citeauthoryear{Je, Lee, Kim, Chan, Yang, and Bianchi}{Je
  et~al\mbox{.}}{2018}]%
        {Pokering}
\bibfield{author}{\bibinfo{person}{Seungwoo Je}, \bibinfo{person}{Minkyeong
  Lee}, \bibinfo{person}{Yoonji Kim}, \bibinfo{person}{Liwei Chan},
  \bibinfo{person}{Xing-Dong Yang}, {and} \bibinfo{person}{Andrea Bianchi}.}
  \bibinfo{year}{2018}\natexlab{}.
\newblock \showarticletitle{PokeRing: Notifications by Poking Around the
  Finger}. In \bibinfo{booktitle}{\emph{Proceedings of the 2018 CHI Conference
  on Human Factors in Computing Systems}} \emph{(\bibinfo{series}{CHI '18})}.
  \bibinfo{publisher}{Association for Computing Machinery},
  \bibinfo{address}{New York, NY, USA}, Article \bibinfo{articleno}{Paper 542},
  \bibinfo{numpages}{10}~pages.
\newblock
\showISBNx{9781450356206}
\urldef\tempurl%
\url{https://doi.org/10.1145/3173574.3174116}
\showDOI{\tempurl}


\bibitem[\protect\citeauthoryear{Je, Rooney, Chan, and Bianchi}{Je
  et~al\mbox{.}}{2017}]%
        {Je2017}
\bibfield{author}{\bibinfo{person}{Seungwoo Je}, \bibinfo{person}{Brendan
  Rooney}, \bibinfo{person}{Liwei Chan}, {and} \bibinfo{person}{Andrea
  Bianchi}.} \bibinfo{year}{2017}\natexlab{}.
\newblock \showarticletitle{tactoRing: A Skin-Drag Discrete Display}. In
  \bibinfo{booktitle}{\emph{Proceedings of the 2017 CHI Conference on Human
  Factors in Computing Systems}}. ACM, \bibinfo{pages}{3106--3114}.
\newblock


\bibitem[\protect\citeauthoryear{Kaaresoja and Linjama}{Kaaresoja and
  Linjama}{2005}]%
        {Kaaresoja2005}
\bibfield{author}{\bibinfo{person}{T. Kaaresoja} {and} \bibinfo{person}{J.
  Linjama}.} \bibinfo{year}{2005}\natexlab{}.
\newblock \showarticletitle{Perception of {Short} {Tactile} {Pulses}
  {Generated} by a {Vibration} {Motor} in a {Mobile} {Phone}}. In
  \bibinfo{booktitle}{\emph{First {Joint} {Eurohaptics} {Conference} and
  {Symposium} on {Haptic} {Interfaces} for {Virtual} {Environment} and
  {Teleoperator} {Systems}}}. \bibinfo{publisher}{IEEE},
  \bibinfo{address}{Pisa, Italy}, \bibinfo{pages}{471--472}.
\newblock
\showISBNx{978-0-7695-2310-1}
\urldef\tempurl%
\url{https://doi.org/10.1109/WHC.2005.103}
\showDOI{\tempurl}


\bibitem[\protect\citeauthoryear{Katayama and Polich}{Katayama and
  Polich}{1998}]%
        {Katayama1998}
\bibfield{author}{\bibinfo{person}{J. Katayama} {and} \bibinfo{person}{J.
  Polich}.} \bibinfo{year}{1998}\natexlab{}.
\newblock \showarticletitle{Stimulus context determines P3a and P3b}.
\newblock \bibinfo{journal}{\emph{Psychophysiology}} \bibinfo{volume}{35},
  \bibinfo{number}{1} (\bibinfo{year}{1998}), \bibinfo{pages}{23--33}.
\newblock


\bibitem[\protect\citeauthoryear{King, Donlin, and Hannaford}{King
  et~al\mbox{.}}{2010}]%
        {King2010}
\bibfield{author}{\bibinfo{person}{H.~H. King}, \bibinfo{person}{R. Donlin},
  {and} \bibinfo{person}{B. Hannaford}.} \bibinfo{year}{2010}\natexlab{}.
\newblock \showarticletitle{Perceptual thresholds for single vs. Multi-Finger
  Haptic interaction}. In \bibinfo{booktitle}{\emph{2010 IEEE Haptics
  Symposium}}. \bibinfo{pages}{95--99}.
\newblock
\showISSN{2324-7347}
\urldef\tempurl%
\url{https://doi.org/10.1109/HAPTIC.2010.5444670}
\showDOI{\tempurl}


\bibitem[\protect\citeauthoryear{Kochhar, Walter J.~Talamonti, and
  Tijerina}{Kochhar et~al\mbox{.}}{2012}]%
        {Kochhar2012}
\bibfield{author}{\bibinfo{person}{Dev~S. Kochhar}, \bibinfo{person}{Jr. Walter
  J.~Talamonti}, {and} \bibinfo{person}{Louis Tijerina}.}
  \bibinfo{year}{2012}\natexlab{}.
\newblock \showarticletitle{Driver Response to Unexpected Automatic
  Braking/Haptic Warning While Backing}.
\newblock \bibinfo{journal}{\emph{Proceedings of the Human Factors and
  Ergonomics Society Annual Meeting}} \bibinfo{volume}{56}, \bibinfo{number}{1}
  (\bibinfo{year}{2012}), \bibinfo{pages}{2211--2215}.
\newblock
\urldef\tempurl%
\url{https://doi.org/10.1177/1071181312561466}
\showDOI{\tempurl}
\showeprint{https://doi.org/10.1177/1071181312561466}


\bibitem[\protect\citeauthoryear{Lakshminarayanan, Lauer, Ramakrishnan,
  Webster, and Seo}{Lakshminarayanan et~al\mbox{.}}{2015}]%
        {Laksh2015}
\bibfield{author}{\bibinfo{person}{Kishor Lakshminarayanan},
  \bibinfo{person}{Abigail~W. Lauer}, \bibinfo{person}{Viswanathan
  Ramakrishnan}, \bibinfo{person}{John~G. Webster}, {and}
  \bibinfo{person}{Na~Jin Seo}.} \bibinfo{year}{2015}\natexlab{}.
\newblock \showarticletitle{Application of vibration to wrist and hand skin
  affects fingertip tactile sensation}.
\newblock \bibinfo{journal}{\emph{Physiological Reports}} \bibinfo{volume}{3},
  \bibinfo{number}{7} (\bibinfo{date}{July} \bibinfo{year}{2015}),
  \bibinfo{pages}{e12465}.
\newblock
\showISSN{2051817X}
\urldef\tempurl%
\url{https://doi.org/10.14814/phy2.12465}
\showDOI{\tempurl}


\bibitem[\protect\citeauthoryear{Lee and Starner}{Lee and Starner}{2010}]%
        {Lee2010}
\bibfield{author}{\bibinfo{person}{Seungyon~"Claire" Lee} {and}
  \bibinfo{person}{Thad Starner}.} \bibinfo{year}{2010}\natexlab{}.
\newblock \showarticletitle{BuzzWear: Alert Perception in Wearable Tactile
  Displays on the Wrist}. In \bibinfo{booktitle}{\emph{Proceedings of the
  SIGCHI Conference on Human Factors in Computing Systems}}
  \emph{(\bibinfo{series}{CHI '10})}. \bibinfo{publisher}{ACM},
  \bibinfo{address}{New York, NY, USA}, \bibinfo{pages}{433--442}.
\newblock
\showISBNx{978-1-60558-929-9}
\urldef\tempurl%
\url{https://doi.org/10.1145/1753326.1753392}
\showDOI{\tempurl}


\bibitem[\protect\citeauthoryear{Ljungberg and Parmentier}{Ljungberg and
  Parmentier}{2012}]%
        {Ljungberg2012}
\bibfield{author}{\bibinfo{person}{Jessica~K. Ljungberg} {and}
  \bibinfo{person}{Fabrice B.~R. Parmentier}.} \bibinfo{year}{2012}\natexlab{}.
\newblock \showarticletitle{Cross-{Modal} {Distraction} by {Deviance}:
  {Functional} {Similarities} {Between} the {Auditory} and {Tactile}
  {Modalities}}.
\newblock \bibinfo{journal}{\emph{Experimental Psychology}}
  \bibinfo{volume}{59}, \bibinfo{number}{6} (\bibinfo{date}{Jan.}
  \bibinfo{year}{2012}), \bibinfo{pages}{355--363}.
\newblock
\showISSN{1618-3169, 2190-5142}
\urldef\tempurl%
\url{https://doi.org/10.1027/1618-3169/a000164}
\showDOI{\tempurl}


\bibitem[\protect\citeauthoryear{Loft, Sanderson, Neal, and Mooij}{Loft
  et~al\mbox{.}}{2007}]%
        {Loft2007}
\bibfield{author}{\bibinfo{person}{Shayne Loft}, \bibinfo{person}{Penelope
  Sanderson}, \bibinfo{person}{Andrew Neal}, {and} \bibinfo{person}{Martijn
  Mooij}.} \bibinfo{year}{2007}\natexlab{}.
\newblock \showarticletitle{Modeling and Predicting Mental Workload in En Route
  Air Traffic Control: Critical Review and Broader Implications}.
\newblock \bibinfo{journal}{\emph{Human Factors}} \bibinfo{volume}{49},
  \bibinfo{number}{3} (\bibinfo{year}{2007}), \bibinfo{pages}{376--399}.
\newblock
\urldef\tempurl%
\url{https://doi.org/10.1518/001872007X197017}
\showDOI{\tempurl}
\showeprint{https://doi.org/10.1518/001872007X197017}
\newblock
\shownote{PMID: 17552304.}


\bibitem[\protect\citeauthoryear{Mars, Debener, Gladwin, Harrison, Haggard,
  Rothwell, and Bestmann}{Mars et~al\mbox{.}}{2008}]%
        {Mars2008}
\bibfield{author}{\bibinfo{person}{R.~B. Mars}, \bibinfo{person}{S. Debener},
  \bibinfo{person}{T.~E. Gladwin}, \bibinfo{person}{L.~M. Harrison},
  \bibinfo{person}{P. Haggard}, \bibinfo{person}{J.~C. Rothwell}, {and}
  \bibinfo{person}{S. Bestmann}.} \bibinfo{year}{2008}\natexlab{}.
\newblock \showarticletitle{Trial-by-{Trial} {Fluctuations} in the
  {Event}-{Related} {Electroencephalogram} {Reflect} {Dynamic} {Changes} in the
  {Degree} of {Surprise}}.
\newblock \bibinfo{journal}{\emph{Journal of Neuroscience}}
  \bibinfo{volume}{28}, \bibinfo{number}{47} (\bibinfo{date}{Nov.}
  \bibinfo{year}{2008}), \bibinfo{pages}{12539--12545}.
\newblock
\showISSN{0270-6474, 1529-2401}
\urldef\tempurl%
\url{https://doi.org/10.1523/JNEUROSCI.2925-08.2008}
\showDOI{\tempurl}


\bibitem[\protect\citeauthoryear{Marsalia, Ferris, Benden, and Zheng}{Marsalia
  et~al\mbox{.}}{2016}]%
        {Marsalia2016}
\bibfield{author}{\bibinfo{person}{Angela~C. Marsalia},
  \bibinfo{person}{Thomas~K. Ferris}, \bibinfo{person}{Mark~E. Benden}, {and}
  \bibinfo{person}{Qi Zheng}.} \bibinfo{year}{2016}\natexlab{}.
\newblock \showarticletitle{Evaluation of {Vibrotactile} {Warning} {Systems}
  for {Supporting} {Hazard} {Awareness} and {Safety} of {Distracted}
  {Pedestrians}}.
\newblock \bibinfo{journal}{\emph{IIE Transactions on Occupational Ergonomics
  and Human Factors}} \bibinfo{volume}{4}, \bibinfo{number}{4}
  (\bibinfo{date}{Oct.} \bibinfo{year}{2016}), \bibinfo{pages}{222--235}.
\newblock
\showISSN{2157-7323, 2157-7331}
\urldef\tempurl%
\url{https://doi.org/10.1080/21577323.2016.1214767}
\showDOI{\tempurl}


\bibitem[\protect\citeauthoryear{McBride, Tran, and Letowski}{McBride
  et~al\mbox{.}}{2017}]%
        {Mcbride2017}
\bibfield{author}{\bibinfo{person}{Maranda McBride}, \bibinfo{person}{Phuong
  Tran}, {and} \bibinfo{person}{Tomasz Letowski}.}
  \bibinfo{year}{2017}\natexlab{}.
\newblock \showarticletitle{Bone {Conduction} {Communication}: {Research}
  {Progress} and {Directions}}.
\newblock  (\bibinfo{year}{2017}), \bibinfo{pages}{148}.
\newblock


\bibitem[\protect\citeauthoryear{Merlo and Hancock}{Merlo and Hancock}{2011}]%
        {Merlo2011}
\bibfield{author}{\bibinfo{person}{James Merlo} {and} \bibinfo{person}{Peter
  Hancock}.} \bibinfo{year}{2011}\natexlab{}.
\newblock \showarticletitle{Quantification of Tactile Cueing for Enhanced
  Target Search Capacity}.
\newblock \bibinfo{journal}{\emph{Military Psychology}} \bibinfo{volume}{23},
  \bibinfo{number}{2} (\bibinfo{year}{2011}), \bibinfo{pages}{137--153}.
\newblock
\urldef\tempurl%
\url{https://doi.org/10.1080/08995605.2011.550226}
\showDOI{\tempurl}
\showeprint{https://doi.org/10.1080/08995605.2011.550226}


\bibitem[\protect\citeauthoryear{Murphy and Dalton}{Murphy and Dalton}{2016}]%
        {Murphy2016}
\bibfield{author}{\bibinfo{person}{Sandra Murphy} {and} \bibinfo{person}{Polly
  Dalton}.} \bibinfo{year}{2016}\natexlab{}.
\newblock \showarticletitle{Out of touch? {Visual} load induces inattentional
  numbness.}
\newblock \bibinfo{journal}{\emph{Journal of Experimental Psychology: Human
  Perception and Performance}} \bibinfo{volume}{42}, \bibinfo{number}{6}
  (\bibinfo{date}{June} \bibinfo{year}{2016}), \bibinfo{pages}{761--765}.
\newblock
\showISSN{1939-1277, 0096-1523}
\urldef\tempurl%
\url{https://doi.org/10.1037/xhp0000218}
\showDOI{\tempurl}


\bibitem[\protect\citeauthoryear{Murphy and Dalton}{Murphy and Dalton}{2018}]%
        {Murphy2018}
\bibfield{author}{\bibinfo{person}{Sandra Murphy} {and} \bibinfo{person}{Polly
  Dalton}.} \bibinfo{year}{2018}\natexlab{}.
\newblock \showarticletitle{Inattentional numbness and the influence of task
  difficulty}.
\newblock \bibinfo{journal}{\emph{Cognition}}  \bibinfo{volume}{178}
  (\bibinfo{date}{Sept.} \bibinfo{year}{2018}), \bibinfo{pages}{1--6}.
\newblock
\showISSN{00100277}
\urldef\tempurl%
\url{https://doi.org/10.1016/j.cognition.2018.05.001}
\showDOI{\tempurl}


\bibitem[\protect\citeauthoryear{Obana, Lim, and Asplund}{Obana
  et~al\mbox{.}}{2020}]%
        {Obana2019}
\bibfield{author}{\bibinfo{person}{T. Obana}, \bibinfo{person}{S.~W.~H. Lim},
  {and} \bibinfo{person}{C.~L. Asplund}.} \bibinfo{year}{2020}\natexlab{}.
\newblock \showarticletitle{Surprise-induced deafness: Unexpected auditory
  stimuli capture attention to the detriment of subsequent detection}.
\newblock \bibinfo{journal}{\emph{psyArXiv}} (\bibinfo{year}{2020}).
\newblock


\bibitem[\protect\citeauthoryear{Parmentier, Ljungberg, Elsley, and
  Lindkvist}{Parmentier et~al\mbox{.}}{2011}]%
        {Parmentier2011}
\bibfield{author}{\bibinfo{person}{Fabrice B.~R. Parmentier},
  \bibinfo{person}{Jessica~K. Ljungberg}, \bibinfo{person}{Jane~V. Elsley},
  {and} \bibinfo{person}{Markus Lindkvist}.} \bibinfo{year}{2011}\natexlab{}.
\newblock \showarticletitle{A behavioral study of distraction by vibrotactile
  novelty.}
\newblock \bibinfo{journal}{\emph{Journal of Experimental Psychology: Human
  Perception and Performance}} \bibinfo{volume}{37}, \bibinfo{number}{4}
  (\bibinfo{year}{2011}), \bibinfo{pages}{1134--1139}.
\newblock
\showISSN{1939-1277, 0096-1523}
\urldef\tempurl%
\url{https://doi.org/10.1037/a0021931}
\showDOI{\tempurl}


\bibitem[\protect\citeauthoryear{Peirce}{Peirce}{2007}]%
        {Peirce2007}
\bibfield{author}{\bibinfo{person}{Jonathan~W. Peirce}.}
  \bibinfo{year}{2007}\natexlab{}.
\newblock \showarticletitle{PsychoPy---Psychophysics software in Python}.
\newblock \bibinfo{journal}{\emph{Journal of Neuroscience Methods}}
  \bibinfo{volume}{162}, \bibinfo{number}{1} (\bibinfo{year}{2007}),
  \bibinfo{pages}{8 -- 13}.
\newblock
\showISSN{0165-0270}
\urldef\tempurl%
\url{https://doi.org/10.1016/j.jneumeth.2006.11.017}
\showDOI{\tempurl}


\bibitem[\protect\citeauthoryear{Qian, Kuber, Sears, and Stanwyck}{Qian
  et~al\mbox{.}}{2014}]%
        {Qian2014}
\bibfield{author}{\bibinfo{person}{Huimin Qian}, \bibinfo{person}{Ravi Kuber},
  \bibinfo{person}{Andrew Sears}, {and} \bibinfo{person}{Elizabeth Stanwyck}.}
  \bibinfo{year}{2014}\natexlab{}.
\newblock \showarticletitle{Determining the Efficacy of Multi-Parameter Tactons
  in the Presence of Real-world and Simulated Audio Distractors}.
\newblock \bibinfo{journal}{\emph{Interacting with Computers}}
  \bibinfo{volume}{26}, \bibinfo{number}{6} (\bibinfo{year}{2014}),
  \bibinfo{pages}{572--594}.
\newblock
\urldef\tempurl%
\url{https://doi.org/10.1093/iwc/iwt054}
\showDOI{\tempurl}


\bibitem[\protect\citeauthoryear{Roumen, Perrault, and Zhao}{Roumen
  et~al\mbox{.}}{2015}]%
        {Roumen2015}
\bibfield{author}{\bibinfo{person}{Thijs Roumen}, \bibinfo{person}{Simon~T.
  Perrault}, {and} \bibinfo{person}{Shengdong Zhao}.}
  \bibinfo{year}{2015}\natexlab{}.
\newblock \showarticletitle{NotiRing: A Comparative Study of Notification
  Channels for Wearable Interactive Rings}. In
  \bibinfo{booktitle}{\emph{Proceedings of the 33rd Annual ACM Conference on
  Human Factors in Computing Systems}} \emph{(\bibinfo{series}{CHI '15})}.
  \bibinfo{publisher}{ACM}, \bibinfo{address}{New York, NY, USA},
  \bibinfo{pages}{2497--2500}.
\newblock
\showISBNx{978-1-4503-3145-6}
\urldef\tempurl%
\url{https://doi.org/10.1145/2702123.2702350}
\showDOI{\tempurl}


\bibitem[\protect\citeauthoryear{Saket, Prasojo, Huang, and Zhao}{Saket
  et~al\mbox{.}}{2013}]%
        {Saket2013}
\bibfield{author}{\bibinfo{person}{Bahador Saket}, \bibinfo{person}{Chrisnawan
  Prasojo}, \bibinfo{person}{Yongfeng Huang}, {and} \bibinfo{person}{Shengdong
  Zhao}.} \bibinfo{year}{2013}\natexlab{}.
\newblock \showarticletitle{Designing an Effective Vibration-based Notification
  Interface for Mobile Phones}. In \bibinfo{booktitle}{\emph{Proceedings of the
  2013 Conference on Computer Supported Cooperative Work}}
  \emph{(\bibinfo{series}{CSCW '13})}. \bibinfo{publisher}{ACM},
  \bibinfo{address}{New York, NY, USA}, \bibinfo{pages}{149--1504}.
\newblock
\showISBNx{978-1-4503-1331-5}
\urldef\tempurl%
\url{https://doi.org/10.1145/2441776.2441946}
\showDOI{\tempurl}


\bibitem[\protect\citeauthoryear{Sathian and Zangaladze}{Sathian and
  Zangaladze}{1996}]%
        {Sathian1996}
\bibfield{author}{\bibinfo{person}{K. Sathian} {and} \bibinfo{person}{A.
  Zangaladze}.} \bibinfo{year}{1996}\natexlab{}.
\newblock \showarticletitle{Tactile spatial acuity at the human fingertip and
  lip}.
\newblock \bibinfo{journal}{\emph{Neurology}} \bibinfo{volume}{46},
  \bibinfo{number}{5} (\bibinfo{year}{1996}), \bibinfo{pages}{1464--1464}.
\newblock
\showISSN{0028-3878}
\urldef\tempurl%
\url{https://doi.org/10.1212/WNL.46.5.1464}
\showDOI{\tempurl}
\showeprint{http://n.neurology.org/content/46/5/1464.full.pdf}


\bibitem[\protect\citeauthoryear{Sergent and Dehaene}{Sergent and
  Dehaene}{2004}]%
        {Sergent2004}
\bibfield{author}{\bibinfo{person}{Claire Sergent} {and}
  \bibinfo{person}{Stanislas Dehaene}.} \bibinfo{year}{2004}\natexlab{}.
\newblock \showarticletitle{Is consciousness a gradual phenomenon? {Evidence}
  for an all-or-none bifurcation during the attentional blink}.
\newblock \bibinfo{journal}{\emph{Psychological Science}} \bibinfo{volume}{15},
  \bibinfo{number}{11} (\bibinfo{year}{2004}), \bibinfo{pages}{720--728}.
\newblock
\urldef\tempurl%
\url{http://pss.sagepub.com/content/15/11/720.short}
\showURL{%
\tempurl}


\bibitem[\protect\citeauthoryear{Shen and Mondor}{Shen and Mondor}{2006}]%
        {Shen2006}
\bibfield{author}{\bibinfo{person}{D. Shen} {and} \bibinfo{person}{T.~A.
  Mondor}.} \bibinfo{year}{2006}\natexlab{}.
\newblock \showarticletitle{Effect of distractor sounds on the auditory
  attentional blink}.
\newblock \bibinfo{journal}{\emph{Perception \& Psychophysics}}
  \bibinfo{volume}{68}, \bibinfo{number}{2} (\bibinfo{year}{2006}),
  \bibinfo{pages}{228--243}.
\newblock


\bibitem[\protect\citeauthoryear{Soto-Faraco, Spence, Fairbank, Kingstone,
  Hillstrom, and Shapiro}{Soto-Faraco et~al\mbox{.}}{2002}]%
        {Soto2002}
\bibfield{author}{\bibinfo{person}{Salvador Soto-Faraco},
  \bibinfo{person}{Charles Spence}, \bibinfo{person}{Katherine Fairbank},
  \bibinfo{person}{Alan Kingstone}, \bibinfo{person}{Anne~P. Hillstrom}, {and}
  \bibinfo{person}{Kimron Shapiro}.} \bibinfo{year}{2002}\natexlab{}.
\newblock \showarticletitle{A crossmodal attentional blink between vision and
  touch}.
\newblock \bibinfo{journal}{\emph{Psychonomic Bulletin \& Review}}
  \bibinfo{volume}{9}, \bibinfo{number}{4} (\bibinfo{date}{Dec.}
  \bibinfo{year}{2002}), \bibinfo{pages}{731--738}.
\newblock
\showISSN{1069-9384, 1531-5320}
\urldef\tempurl%
\url{https://doi.org/10.3758/BF03196328}
\showDOI{\tempurl}


\bibitem[\protect\citeauthoryear{Squires, Squires, and Hillyard}{Squires
  et~al\mbox{.}}{1975}]%
        {Squires1975}
\bibfield{author}{\bibinfo{person}{Nancy~K Squires}, \bibinfo{person}{Kenneth~C
  Squires}, {and} \bibinfo{person}{Steven~A Hillyard}.}
  \bibinfo{year}{1975}\natexlab{}.
\newblock \showarticletitle{Two varieties of long-latency positive waves evoked
  by unpredictable auditory stimuli in man}.
\newblock \bibinfo{journal}{\emph{Electroencephalography and Clinical
  Neurophysiology}} \bibinfo{volume}{38}, \bibinfo{number}{4}
  (\bibinfo{date}{April} \bibinfo{year}{1975}), \bibinfo{pages}{387--401}.
\newblock
\showISSN{00134694}
\urldef\tempurl%
\url{https://doi.org/10.1016/0013-4694(75)90263-1}
\showDOI{\tempurl}


\bibitem[\protect\citeauthoryear{Stanley}{Stanley}{2006}]%
        {Stanley2006}
\bibfield{author}{\bibinfo{person}{Laura~M. Stanley}.}
  \bibinfo{year}{2006}\natexlab{}.
\newblock \showarticletitle{Haptic and Auditory Cues for Lane Departure
  Warnings}.
\newblock \bibinfo{journal}{\emph{Proceedings of the Human Factors and
  Ergonomics Society Annual Meeting}} \bibinfo{volume}{50},
  \bibinfo{number}{22} (\bibinfo{year}{2006}), \bibinfo{pages}{2405--2408}.
\newblock
\urldef\tempurl%
\url{https://doi.org/10.1177/154193120605002212}
\showDOI{\tempurl}
\showeprint{https://doi.org/10.1177/154193120605002212}


\bibitem[\protect\citeauthoryear{Vachon, Labont{\'e}, and Marsh}{Vachon
  et~al\mbox{.}}{2017}]%
        {Vachon2017}
\bibfield{author}{\bibinfo{person}{Fran{\c c}ois Vachon},
  \bibinfo{person}{Katherine Labont{\'e}}, {and} \bibinfo{person}{John~E.
  Marsh}.} \bibinfo{year}{2017}\natexlab{}.
\newblock \showarticletitle{Attentional capture by deviant sounds: {A}
  noncontingent form of auditory distraction?}
\newblock \bibinfo{journal}{\emph{Journal of Experimental Psychology: Learning,
  Memory, and Cognition}} \bibinfo{volume}{43}, \bibinfo{number}{4}
  (\bibinfo{year}{2017}), \bibinfo{pages}{622--634}.
\newblock
\showISSN{1939-1285, 0278-7393}
\urldef\tempurl%
\url{https://doi.org/10.1037/xlm0000330}
\showDOI{\tempurl}


\bibitem[\protect\citeauthoryear{Vega-Bermudez and Johnson}{Vega-Bermudez and
  Johnson}{2001}]%
        {VegaBermudez2001}
\bibfield{author}{\bibinfo{person}{Francisco Vega-Bermudez} {and}
  \bibinfo{person}{Kenneth~O. Johnson}.} \bibinfo{year}{2001}\natexlab{}.
\newblock \showarticletitle{Differences in spatial acuity between digits}.
\newblock \bibinfo{journal}{\emph{Neurology}} \bibinfo{volume}{56},
  \bibinfo{number}{10} (\bibinfo{date}{May} \bibinfo{year}{2001}),
  \bibinfo{pages}{1389}.
\newblock
\urldef\tempurl%
\url{https://doi.org/10.1212/WNL.56.10.1389}
\showDOI{\tempurl}


\bibitem[\protect\citeauthoryear{Weinstein}{Weinstein}{1962}]%
        {Weinstein1962}
\bibfield{author}{\bibinfo{person}{S. Weinstein}.}
  \bibinfo{year}{1962}\natexlab{}.
\newblock \showarticletitle{Tactile sensitivity of the phalanges}.
\newblock \bibinfo{journal}{\emph{Perceptual and Motor Skills}}
  \bibinfo{volume}{14} (\bibinfo{year}{1962}), \bibinfo{pages}{351--354}.
\newblock


\bibitem[\protect\citeauthoryear{Weisenberger and Craig}{Weisenberger and
  Craig}{1982}]%
        {Weisenberger1982}
\bibfield{author}{\bibinfo{person}{Janet~M. Weisenberger} {and}
  \bibinfo{person}{James~C. Craig}.} \bibinfo{year}{1982}\natexlab{}.
\newblock \showarticletitle{A tactile metacontrast effect}.
\newblock \bibinfo{journal}{\emph{Perception \& Psychophysics}}
  \bibinfo{volume}{31}, \bibinfo{number}{6} (\bibinfo{date}{Nov.}
  \bibinfo{year}{1982}), \bibinfo{pages}{530--536}.
\newblock
\showISSN{0031-5117, 1532-5962}
\urldef\tempurl%
\url{https://doi.org/10.3758/BF03204185}
\showDOI{\tempurl}


\bibitem[\protect\citeauthoryear{Yamaguchi and Knight}{Yamaguchi and
  Knight}{1991}]%
        {Yamaguchi1991}
\bibfield{author}{\bibinfo{person}{Shuhei Yamaguchi} {and}
  \bibinfo{person}{Robert~T. Knight}.} \bibinfo{year}{1991}\natexlab{}.
\newblock \showarticletitle{P300 generation by novel somatosensory stimuli}.
\newblock \bibinfo{journal}{\emph{Electroencephalography and Clinical
  Neurophysiology}} \bibinfo{volume}{78}, \bibinfo{number}{1}
  (\bibinfo{date}{Jan.} \bibinfo{year}{1991}), \bibinfo{pages}{50--55}.
\newblock
\showISSN{00134694}
\urldef\tempurl%
\url{https://doi.org/10.1016/0013-4694(91)90018-Y}
\showDOI{\tempurl}


\bibitem[\protect\citeauthoryear{Yatani, Gergle, and Truong}{Yatani
  et~al\mbox{.}}{2012}]%
        {Yatani2012}
\bibfield{author}{\bibinfo{person}{Koji Yatani}, \bibinfo{person}{Darren
  Gergle}, {and} \bibinfo{person}{Khai Truong}.}
  \bibinfo{year}{2012}\natexlab{}.
\newblock \showarticletitle{Investigating effects of visual and tactile
  feedback on spatial coordination in collaborative handheld systems}. In
  \bibinfo{booktitle}{\emph{Proceedings of the ACM 2012 conference on Computer
  Supported Cooperative Work}}. ACM, \bibinfo{pages}{661--670}.
\newblock


\bibitem[\protect\citeauthoryear{Yatani and Truong}{Yatani and Truong}{2009}]%
        {Yatani2009}
\bibfield{author}{\bibinfo{person}{Koji Yatani} {and}
  \bibinfo{person}{Khai~Nhut Truong}.} \bibinfo{year}{2009}\natexlab{}.
\newblock \showarticletitle{SemFeel: a user interface with semantic tactile
  feedback for mobile touch-screen devices}. In
  \bibinfo{booktitle}{\emph{Proceedings of the 22nd annual ACM symposium on
  User interface software and technology}}. ACM, \bibinfo{pages}{111--120}.
\newblock


\bibitem[\protect\citeauthoryear{{Ying Zheng}, Su, and Morrell}{{Ying Zheng}
  et~al\mbox{.}}{2013}]%
        {Zheng2013}
\bibfield{author}{\bibinfo{person}{{Ying Zheng}}, \bibinfo{person}{E. Su},
  {and} \bibinfo{person}{J.~B. Morrell}.} \bibinfo{year}{2013}\natexlab{}.
\newblock \showarticletitle{Design and evaluation of pactors for managing
  attention capture}. In \bibinfo{booktitle}{\emph{2013 {World} {Haptics}
  {Conference} ({WHC})}}. \bibinfo{publisher}{IEEE},
  \bibinfo{address}{Daejeon}, \bibinfo{pages}{497--502}.
\newblock
\showISBNx{978-1-4799-0088-6 978-1-4799-0087-9}
\urldef\tempurl%
\url{https://doi.org/10.1109/WHC.2013.6548458}
\showDOI{\tempurl}


\end{thebibliography}

\end{document}